\documentclass[lettersize,journal]{IEEEtran}
\usepackage{amsmath,amsfonts}
\usepackage{algorithmic}
\usepackage{algorithm}
\usepackage{array}
\usepackage[caption=false,font=normalsize,labelfont=sf,textfont=sf]{subfig}
\usepackage{textcomp}
\usepackage{stfloats}
\usepackage{url}
\usepackage{verbatim}
\usepackage{graphicx}
\usepackage{cite}
\usepackage[nolist]{acronym}
\usepackage{tabularx}
\usepackage{url}
\usepackage{siunitx}
\usepackage{cuted}\stripsep -3pt plus 3pt minus 2pt
\begin{document}
	\title{Communication, Sensing and Control integrated Closed-loop System: Modeling, Control Design and Resource Allocation}
	\author{		
		Zeyang Meng, Dingyou Ma, Zhiqing Wei, \IEEEmembership{Member, IEEE},\\  Ying Zhou, Zhiyong Feng, \IEEEmembership{Senior Member, IEEE}
		\thanks{
			This work was supported in part by the National Key Research and Development Program of China under Grant 2020YFA0711302, in part by the National Natural Science Foundation of China (NSFC) under Grant 92267202.
			
			Parts of this work were presented at the 2023 IEEE Global Communications Conference as \cite{ConferencePaper}.
			
			Zeyang Meng, Dingyou Ma, Zhiqing Wei, Ying Zhou and Zhiyong Feng are with Key Laboratory of the Universal Wireless Communications, Beijing University of Posts and Telecommunications, Beijing, China 100876 (email: \{mengzeyang, dingyouma, weizhiqing, zhouying9705, fengzy\}@bupt.edu.cn

			Correspondence authors: Zhiyong Feng.}
	}
	
	\maketitle
	
	\begin{abstract}
		The wireless communication technologies have fundamentally revolutionized industrial operations. The operation of the automated equipment is conducted in a closed-loop manner, where the status of devices is collected and sent to the control center through the uplink channel, and the control center sends the calculated control commands back to the devices via downlink communication. However, existing studies neglect the interdependent relationship between uplink and downlink communications, and there is an absence of a unified approach to model the communication, sensing, and control within the loop. This can lead to inaccurate performance assessments, ultimately hindering the ability to provide guidance for the design of practical systems. Therefore, this paper introduces an integrated closed-loop model that encompasses sensing, communication, and control functionalities, while addressing the coupling effects between uplink and downlink communications. Through the analysis of system convergence, an inequality pertaining to the performances of sensing, communication, and control is derived. Additionally, a joint optimization algorithm for control and resource allocation is proposed. Simulation results are presented to offer an intuitive understanding of the impact of system parameters. The findings of this paper unveil the intricate correlation among sensing, communication, and control, providing insights for the optimal design of industrial closed-loop systems.
	\end{abstract}
	
	\begin{IEEEkeywords}
		closed-loop system, wireless network, effective-capacity, model predictive control
	\end{IEEEkeywords}
	
	\section{Introduction}
	
	Recently, the rapid progress of wireless communication technologies, particularly 5G and the forthcoming 6G, has initiated a fundamental change in industrial settings \cite{5G_standard_22261, 6G_WhitePaper}. 
	The wireless network liberate automated equipment from the constraints of cables, ensuring mobility for devices such as the mobile robotic arm and \ac{AGV}.
	This advancement enables manufacturers to flexibly alter the structure of production line and adjust manufacturing processes based on the order demands, thereby enhancing the production efficiency and reducing manufacturing costs \cite{5G_Industry_1, 5G_Industry_2}.
	
\begin{figure}
		\centering
		\includegraphics[width=0.35\textheight]{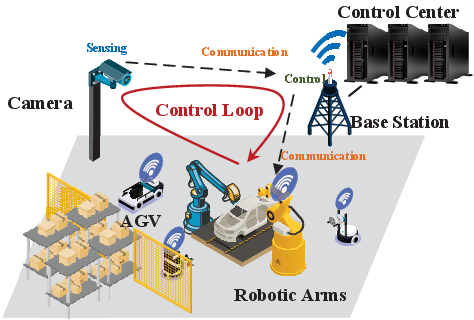}
		\DeclareGraphicsExtensions.
		\caption{A typical closed-loop process in the typical industry with wireless network.}	
		\label{Scenario}
	\end{figure}

	The operational process of automated equipment with the wireless network is based on the iterative closed-loop process, as shown in Fig. \ref{Scenario}.
	Specifically, sensors collect the status information of equipment, which is then transmitted to the control center via uplink wireless channels. 
	The control center generate the control commands based on the sensed data and then sends these commands back to the controlled equipment through downlink wireless communication.
	In such a process, the performance of communication, sensing and control are closely interrelated. 
	Sensing affects not only the accuracy of control commands but also the data load in the communication process. 
	In addition, the communication capability of the system determines whether the control center can receive fresh sensing information in a timely manner.
	Therefore, the modeling of the closed-loop process and the associated performance is a challenging and complex task.
	
	Currently, existing studies have examined such systems, which are mainly classified into two kinds of approaches: control-centric and communication-centric.
	Control-centric approaches are generally based on the dynamic equations of devices under simple conditions.
	Specifically, communication performance is simplified as predefined constants \cite{ConstantDelay_1, ConstantDelay_2, ConstantDelay_3} or bounded random variables \cite{RandomDelay_1, RandomDelay_2}, and is then integrated into dynamic equations.
	Sensing performance, including quantization errors \cite{Quantization_1} and estimation errors \cite{Estimation_1}, is also taken into consideration in some research.
	Subsequently, the stability of the system is analyzed, and new robust control methods are proposed based on the new dynamic equations.
	The limitation of this approach lies in the disparity between theoretical communication models and real-world communication systems.
	This gap hinders the achievement of expected control effects and poses challenges in the design of communication systems.
	In addition, there is a lack of unified modeling of interactions among communication, sensing, and control.

	Communication-centric approaches employ parameters such as \ac{LQR} cost \cite{ComLQR_1, ComLQR_2} and convergence rate \cite{ComConvergence_1, ComConvergence_2} to model the efficiency of control systems. 
	These parameters are integrated with the metrics of communication performance, such as mutual information \cite{ComLQR_1,ComLQR_2} and channel capacity \cite{ComCapacity_1, ComCapacity_2}, to establish a unified performance index or to derive an optimal design for the communication system.
	The limitation of this approach is that the uplink and downlink communications are modeled independently, which contradicts the characteristic of control loops where the data carried by the downlink is closely related to the uplink sensing data. 
	Additionally, preset control parameters make it difficult for the system to achieve globally optimal control effects.
	Furthermore, few studies consider the impact of sensing errors on system performance, which may lead to deviations in performance when communication designs are applied in practical systems.
	
	To address the above shortcomings, we establish a sensing-communication-control integrated closed-loop model to address the aforementioned shortcomings, and propose a joint optimization method for control and resource allocation. 
	Specifically, to address the issue of the separation between uplink and downlink models of communication-centric approaches, we develop an uplink-downlink coupled communication model based on the effective capacity theory, which establishes a correlation between the network performance and communication resources. 
	On this basis, closed-form expressions for closed-loop delay and packet loss rate are derived to quantitatively describe the key communication indicators that influence control performances, thereby resolves the problem of over simplified communication metrics in control-centric approaches.
	
	Furthermore, in order to describe the interaction of communication, sensing, and control processes, we develop a control model that accounts for delay, packet loss and estimation error. Additionally, we formulate a sensing-estimation model to derive the boundary of the estimation error.
	To clarify the complex relationship among communication, sensing, and control parameters, an inequality involving convergence rate, bandwidth, and quantization level is derived by applying Lyapunov stability theory \cite{LyapunovTheoryIntro} based on the proposed model.

	Finally, to provide guidance for both communication and control within the closed-loop system, a joint optimization problem for control and resource allocation is proposed, which is highly non-convex.
	To address this optimization challenge, a \ac{DE}-based optimization algorithm \cite{DEIntro} is employed to acquire global optimum solutions.
	The simulation demonstrates the nonlinear effect of parameters such as closed-loop delay, convergence rate, and quantization level, whose excessively high or low values will adversely impact the accomplishment of control tasks due to the coupling relationship among communication, sensing, and control.
	These phenomena have not yet been explored in existing research.
	
	The main contributions are summarized as follows. 
	\begin{itemize}
		\item Based on the effective capacity theory, a joint modeling of the uplink and downlink communication processes is conducted. From this model, closed-form expressions for closed-loop delay and packet loss rate are derived, establishing a relationship between the closed-loop performance and network resources.
		\item An inequality for the system convergence rate is derived by incorporating the effects of three aspects: sensing, communication, and control, which characterizes the complex constraining relationship among these three functionalities.
		\item A joint optimization algorithm is proposed, achieving efficient utilization of communication resources while ensuring  control effectiveness. The global optimal solution for this problem is obtained based on a heuristic method.
	\end{itemize}
	
	The rest of this article is organized as follows. 
	In Section \uppercase\expandafter{\romannumeral2}, the system model of the wireless closed-loop control system is established, which provides the main analytical results of this article.
	In Section \uppercase\expandafter{\romannumeral3}, the joint optimization of sensing, control and communication is carried out, where a heuristic method is applied to solve the non-convex problem.
	In Section~\uppercase\expandafter{\romannumeral4}, simulation results are provided for both the result of the optimization problem and the control strategy proposed in the system model.
	Finally, concluding remarks are provided in Section \uppercase\expandafter{\romannumeral5}.
	The nomenclature of this article is shown in Table~\ref{Nomenclature}.

	\begin{table*}[!h]
		\caption{Nomenclature}
		\renewcommand{\arraystretch}{1.3} 
		\label{Nomenclature}
		\begin{center}
			
			\begin{tabular}{|m{0.1\textwidth}|m{0.35\textwidth}|m{0.1\textwidth}|m{0.35\textwidth}|}
				\hline
				\textbf{Symbol} & \textbf{Meaning} & \textbf{Symbol} & \textbf{Meaning} \\
				\hline
				$\lambda_u / \lambda_d$ & Arrival rate of the uplink/downlink queue. 
				& $L_u$ & Departure rate of the uplink queue.\\
				\hline
				$R_u / R_d$ & The capacity of uplink/downlink channels.  
				&$W_u/W_d$ & The allocated bandwidth of uplink/downlink channels.\\
				\hline
				$\gamma$ & The fading coefficient of the channel.
				& $\beta_u/\beta_d$ & The parameter of exponential distributed random variable $\gamma^2$.\\
				\hline
				$C_u / C_d$ & The effective capacity of the uplink/downlink queue.
				& $\theta_u/\theta_d$ & The decay rate of queue overflow probability.\\
				\hline
				$D_c$ & The closed-loop delay.
				&$D_{c,\max}$ & The threshold of $D_c$ for packet loss.\\
				\hline
				$\epsilon_c$ & The packet loss rate.
				& $T_d$ & The time interval of the discretized state.\\
				\hline
				$\mathbf{X}$ & The state of the device.
				&$\hat{\mathbf{X}}$ & The estimation of the state.\\
				\hline
				$\mathbf{K}$ & The control law coefficient.
				& $\eta$& The random variable corresponding to packet loss.\\
				\hline
				$\mathbf{X}_L/\mathbf{X}_U$ & The bounds of the state.
				&$r$& The quantization level.\\
				\hline
				$\mathbf{e}_\tau$ & The estimation error at time $\tau$ 
				&$\rho$ & The convergence rate of the system.\\
				\hline

			\end{tabular}
		\end{center}
	\end{table*}

	\section{System Model}
	\begin{figure}
		\centering
		\includegraphics[width=0.35\textheight]{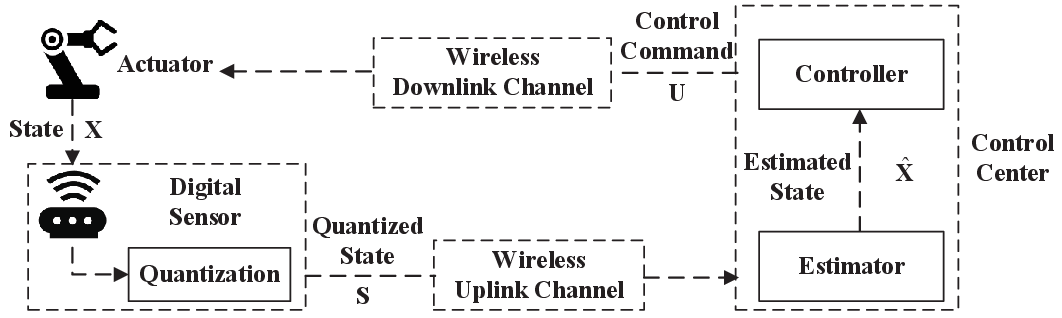}
		\DeclareGraphicsExtensions.
		\caption{System model of a wireless closed-loop control system.}	
		\label{introductionFig}
	\end{figure}
	The system model of a wireless closed-loop system is shown in Fig. \ref{introductionFig}, which is composed of the actuator, the digital sensor, two wireless channels and a control center \cite{SystemModel_1, SystemModel_2, SystemModel_3}.
	The closed-loop process starts from the digital sensor, which perceives the state information of the actuator and quantifies it into quantized values.
	These quantized states are then transmitted to the control center through the wireless uplink channel, and processed to acquire the estimated state to recover the state of the actuator.
	Subsequently, the controller at the control center generates control commands with the estimated state and transmit them to the actuator through the wireless downlink channel.
	The commands are then executed on the actuator, so that the closed-loop process are completed.
	
	In the closed-loop process, it is evident that sensing, communication, and control interact with each other. 
	To better achieve the closed-loop control of devices, it is necessary to construct a closed-loop control model.
	In the sequel, the system model is introduced separately from the perspectives of communication, control, and sensing. 
	The coupling inequality among the performances of these three functions is derived from the viewpoint of system convergence.

	\subsection{Communication Model}
	In a closed-loop system, the wireless communication primarily handles the uploading of sensing data and the dispatching of control instructions.
	The communication models of related researches mainly focus on the transmission delay in either uplink or downlink transmission, specifically the duration required for a packet to transmit from the first bit to the last bit.
	This approach, however, has the following two shortcomings.
	On the one hand, existing studies overlook the coupled characteristics between uplink and downlink transmissions.
	In closed-loop systems, the transmission scale of the downlink data is influenced by the volume of uplink packets and their arrival interval, since the control data transmitted by the downlink channel is generated by the sensing data from the uplink channel. 
	Therefore, a model that treats uplink and downlink transmissions independently is not consistent with reality.
	On the other hand, the impact of queuing delays of the packets have been disregarded.
	Compared with transmission delay, queuing delay constitutes a more significant portion of the delay components in industrial networks where data packets are frequently transmitted, leading to great randomness in the transmission link \cite{DelayComponent_1, DelayComponent_2}.
	
	Therefore, a closed-loop communication model is constructed with uplink-downlink tandem queues.
	The performance of the communication system is measured by three typical indicators, i.e. the closed-loop delay $D_c$, the packet loss rate $\epsilon_c$, and the maximum arrival rate $\lambda_{\max}$.
	The closed-loop delay $D_c$ represents the total time from the transmission of sensing packets to the reception of the corresponding control instructions. 
	Excessive closed-loop delay can lead to imprecise control instructions, thereby affecting the control effectiveness.
	The packet loss rate $\epsilon_c$ reflects the probability of data packets being discarded due to factors such as timeouts during the entire communication process, which results in reduced system robustness and lower convergence ability.
	Besides, the maximum arrival rate $\lambda_{\max}$ indicates the maximum amount of data that the system can handle in a unite of time, which is affected by factors such as bandwidth and channel conditions.
	In the closed-loop system, $\lambda_{\max}$ is the maximum amount of sensing data that can be transmitted per second in the uplink channel.

	In the following analysis, we first establish a model of a single queue.
	The closed-form expression for the effective capacity of a single queue under Rayleigh fading channels, as well as the limitations of the arrival rate, is derived.
	On the basis of these results, the properties of the closed-loop delay, the packet loss rate, and the maximum arrival rate in tandem queues are derived, and the closed-form results are presented to better assist in system design.

	\subsubsection{The Single Queue Effective Capacity Model}
	The derivation of this part is based on the following assumptions.
	
	\textbf{\textit{Assumption 1: }}The arrival of data packets in the uplink queue is periodic and the size of each packet is the same.
	
	\textbf{\textit{Assumption 2: }} 
	The size of the control data packets is consistent, and the amount of sensing data required to generate each control command is fixed.
	
	\textbf{\textit{Assumption 3: }}The channel fading follows a Rayleigh distribution.

	Assumption 1 holds because the packets in the uplink queue originate from sensors, which usually sense periodically in the factory, and the format of these packets is typically consistent.
	Assumption 2 is determined by the control algorithm. 
	Since the control algorithm often remains unchanged in each iteration, it is reasonable to consider Assumption 2 as valid.
	The Assumption 3 is reasonable because, in factories, there are often many obstructions and scatters, with rare direct paths.
	
	Under the above assumptions, the number of arrival data per second, i.e. the arrival rate, of the uplink queue $\lambda_u$ \footnote{For brevity, we utilize the subscript "u" to denote "uplink" and the subscript "d" to signify "downlink" in the rest of this article.} is constant according to Assumption 1.
	Besides, according to Assumption~2, the arrival rate of the downlink queue, denoted by $\lambda_d$, has a linear relationship with the amount of data departing from the uplink queue per second, i.e. the uplink departure rate $L_u$, which is 
	\begin{equation}\label{lambdaD}
		\lambda_d = c_d L_u,
	\end{equation}
	with $c_d$ being a constant.
	
	Besides, the service rate of the queue denotes the average number of items that can be served by the queue per unit of time.
	In the proposed communication model, the service rate is the average number of packages transmitted per second, i.e. the transmission rate of the communication channel.
	Therefore, according to Assumption 3, the service rate of two queues, i.e. the capacity of uplink and downlink channels $R_u$ and $R_d$ satisfies
	\begin{equation}
		\label{Ri}
		R_i = W_i \log_2(1+\text{SNR}_i\cdot\gamma_i^2)  \quad (i=u,d),
	\end{equation}
	where $\text{SNR}_i$ is a constant obtained by dividing the expectation of the signal power by the expectation of the noise power, and $W_i$ is the bandwidth of the uplink and downlink channels. Parameter $\gamma$ is the fading coefficient, which is a random variable following the Rayleigh distribution. Therefore, $\gamma^2$ follows the exponential distribution \cite{RayExp}. 
	
	Subsequently, the effective capacity theory is employed to connect the link-layer performance with physical layer resources, where the \emph{effective capacity} is defined as the maximum acceptable amount of bits that the channel can handle per unit of time, which is given by \cite{EffectiveCapacity},
	\begin{equation}
		\begin{aligned}
			\label{EC}
			C_i(\theta_i, R_i) = &-\frac{1}{\theta_i}\ln\left(\mathbb{E}\{\exp(-\theta_i R_i)\}\right) \quad (i=u,d),
		\end{aligned} 
	\end{equation}
	where $C_u$ and $C_d$ are effective capacities of the uplink and downlink communications, respectively. The parameter $\theta_i$ represents the decay rate of queue overflow probability, which satisfies
	\begin{equation}
		\lim _{q_{i,0} \rightarrow \infty} \frac{\ln \operatorname{Pr}\{q_i\left(\infty\right) \geq q_{i,0}\}}{q_{i,0}} = -\theta_i \quad \left(i=u,d\right) ,
	\end{equation}
	with $q_i\left( \infty \right)$ being the length of the communication buffer queue in steady state, and $q_{i,0}$ being the buffer overflow threshold.
	
	With Assumption 3, the effective capacities of the uplink and downlink queues follow Theorem \ref{ECSingleQueue}.
	\newtheorem{theorem}{Theorem}
	\begin{theorem}[Effective Capacity of Rayleigh Channel] \label{ECSingleQueue}
		The effective capacity of uplink and downlink queues under Assumption 3 is approximated by
		\begin{equation}
			\scalebox{1}{$
				\begin{aligned}
					C_i&(\theta_i, W_i, \text{SNR}_i,\beta_i) = \\
					&\frac{1}{\theta_i} \ln (\text{SNR}_i \beta_i)(\frac{W_i \theta_i}{\ln 2}-1)-\frac{1}{\text{SNR}_i \beta_i}\quad (i=u,d),
				\end{aligned}
				$}
		\end{equation}
		where $\beta_i$ is the rate parameter of the random variable $\gamma^2$ in the uplink and downlink channels.
	\end{theorem}
	\vspace{0.2cm}
	\begin{IEEEproof}
		See the proof in Appendix A.
	\end{IEEEproof}
	
	According to the definition of the effective capacity, $C_i$ is the maximum data volume that the queue can accommodate. Therefore, it should satisfy the following inequalities
	\begin{equation}\label{EAleEC}
		\scalebox{1}{$
			\begin{aligned}
				&\lambda_i \le C_i(\theta_i, W_i, \text{SNR}_i,\beta_i)  \\
				&=\frac{1}{\theta_i} \ln (\text{SNR}_i \beta_i)\left(\frac{W_i \theta_i}{\ln 2}-1\right)-\frac{1}{\text{SNR}_i \beta_i}\quad (i=u,d).
			\end{aligned}
			$}
	\end{equation}
	
	Moreover, packets are considered to be lost when their delay exceeds a specified threshold. 
	Consequently, the packet loss rate, denoted by $\epsilon_i$, is defined as
	\begin{equation}\label{singleQueueDelay}
		\scalebox{1}{$
			\begin{aligned}
				\epsilon_i = \mathrm{P}\left(D_i>D_{i,\max }\right)=e^{-D_{i,\max } \theta_i C_i(\theta_i, W_i, \text{SNR}_i,\beta_i)},
			\end{aligned}
			$}
	\end{equation}
	where $D_i$ is the delay of the packet, and $D_{i,\max }$ is the threshold for packet loss\cite[Eqn. 3]{ECSimpEq} .
	
	\subsubsection{The Analysis of the Tandem Queue}
	Based on the analysis of a single queue above, we further consider the performance metrics for the tandem queue of the close-loop communication.
	The closed-form expressions of the closed-loop delay $D_c$, the packet loss rate $\epsilon_c$, and the maximum arrival rate $\lambda_{\max}$ are given in the following theorems.
	
	\begin{theorem}[Packet Loss Rate and Closed-loop Delay] \label{epsiloncAndDc}
		Considering that $D_{c, \max}$ is the threshold of the closed-loop delay $D_c$ that leads to packet loss, the packet loss rate $\epsilon_c$ is given by 
		\begin{equation} \label{epsilon_cResult}
			\scalebox{1}{$
				\begin{aligned}
					\begin{aligned}
						\epsilon_c=\frac{\mathrm{e}^{-D_{c, \operatorname{max}} \mu_d} \mu_u-\mathrm{e}^{-D_{c, \max } \mu_u} \mu_d}{\mu_u-\mu_d}.
					\end{aligned}
				\end{aligned}
				$}
		\end{equation}
		The exception of the closed-loop delay is derived as	
		\begin{equation} \label{EDc}
			\scalebox{1}{$
				\begin{aligned}
					\begin{aligned}
						&E\left[D_c|D_c<D_{c,\max}\right]\\
						=&[\mu_u^2-\mu_d^2+\mathrm{e}^{-D_{c, \max } \mu_u} \mu_d^2\left(1+D_{c, \max } \mu_u\right)\\
						&\ \ -\mathrm{e}^{-D_{c, \max x} \mu_d} \mu_u^2\left(1+D_{c, \max } \mu_d\right)] /\\
						&\mu_u \mu_d\left(\mu_u-\mu_d+\mathrm{e}^{-D_{c, \max x} \mu_u} \mu_d-\mathrm{e}^{-D_{c, \max } \mu_d} \mu_u\right),
					\end{aligned}
				\end{aligned}
				$}
		\end{equation}
		where
		\begin{equation} 
			\scalebox{1}{$
				\begin{aligned}
					\mu_i =\ln (\text{SNR}_i \beta_i)(\frac{W_i \theta_i}{\ln 2}-1)-\frac{\theta_i}{\text{SNR}_i \beta_i}\ \ (i=u,d).
				\end{aligned}
				$}
		\end{equation}
	\end{theorem}
	\vspace{0.2cm}
	\begin{IEEEproof}
		See the proof in Appendix B.
	\end{IEEEproof}

	\begin{theorem}[The Maximum Arrival Rate] \label{arrivalRateLimit}
		When $\theta_u > \theta_d$, the maximum arrival rate $\lambda_{\max}$ satisfy 
		\begin{equation} \label{lambdaIneq1}
			\scalebox{1}{$
				\begin{aligned}
					&\lambda_{\max} \\
					=&\min \Bigg\{\frac{1}{\theta_u}\ln (\text{SNR}_u \beta_u)(\frac{W_u \theta_u}{\ln 2}-1)-\frac{1}{\text{SNR}_u \beta_u}, \\
					&\quad\quad\frac{1}{c_d \theta_d} \ln (\text{SNR}_d \beta_d)(\frac{W_d \theta_d}{\ln 2}-1)-\frac{1}{\text{SNR}_d \beta_d} \Bigg\}.
				\end{aligned}
				$}
		\end{equation}
		
		Besides, when $\theta_u < \theta_d$,
		
		\begin{equation} \label{lambdaIneq2}
			\scalebox{1}{$
				\begin{aligned}
					&\lambda_{\max} \\
					=& \min \Bigg\{ 
					\begin{aligned}[t]
						&\frac{1}{\theta_u}\ln (\text{SNR}_u \beta_u)\left(\frac{W_u \theta_u}{\ln 2}-1\right)-\frac{1}{\text{SNR}_u \beta_u}, \\
						&\frac{1}{c_d \theta_u} \ln \Bigg[\left(\text{SNR}_d \beta_d\right)\left(\frac{W_d \theta_d}{\ln 2}\right)\left(\text{SNR}_u \beta_u\right)^{c_d}\\
						&\qquad \qquad \left(\frac{W_u\left(\theta_d-\theta_u\right)}{\ln 2}\right)^{c_d}\Bigg]\Bigg\}.
					\end{aligned}
				\end{aligned}
				$}
		\end{equation}

	\end{theorem}
	\vspace{0.2cm}
	\begin{IEEEproof}
		See the proof in Appendix C.
	\end{IEEEproof}

	\subsection{Control Model}\label{ControlModel}
	The control system serves as the backbone of a closed-loop system. 
	The purpose of the control stage is to generate the control commands based on the current state of the device, so as to enable the device to reach the expected state after a period of time.
	In the field of control, dynamic functions are often applied to model the state evolution.
	The following subsections progressively establish the dynamic function of the closed-loop system, 
	advancing from a simple model to a model with imperfect sensing, and finally to a model influenced by imperfect wireless communication.

	
	\subsubsection{The Basic Control Model}
	The model starts from a basic state-space control model \cite{ControlFunction}, i.e.
	\begin{equation} \label{basicControlModel}
		\scalebox{1}{$
			\begin{aligned}
				\dot{\mathbf{X}}=\mathbf{A} \mathbf{X}+\mathbf{B} \mathbf{U},
			\end{aligned}
			$}
	\end{equation}
	where $\mathbf{X} = \{x_1,x_2,\dots,x_n\}$ is the state vector of the actuator \textcolor{black}{with $n$ states}, $\mathbf{U}$ is the control vector, which is determined by the control command. 
	$\mathbf{A}$ and $\mathbf{B}$ are system matrices which are guaranteed by physical characteristics of the system.
	The state evolution of nearly all devices can be represented in this form, such as \acp{AGV} \cite{AGVControlModel} and robotic arms \cite{ArmControlModel}.
	
	For example, for the movement control of \acp{AGV} \cite{AGVControlModel}, $\mathbf{X}=\left[\delta, v, a\right]^T$, with $\delta$, $v$, and $a$ being the position, velocity, and acceleration, respectively. $\mathbf{U}$ is a linear control strategy, which is the linear combination of the state $\mathbf{X}$. With such control strategy, 
	$\mathbf{A}$ and $\mathbf{B}$ are given by
	\begin{equation}
		\mathbf{A}=\left[\begin{array}{ccc}0 & 1 & 0 \\ 0 & 0 & 1 \\ 0 & 0 & -1 / \varsigma\end{array}\right],
	\end{equation}
	and 
	\begin{equation}
		\mathbf{B}=\left[\begin{array}{c}0 , 0 , -1 / \varsigma\end{array}\right]^\mathrm{T},
	\end{equation}
	respectively, where $\varsigma$ being a constant related to the engine.
	
	For ease of analysis, \eqref{basicControlModel} can be discretized with the Euler's method \cite{Euler} by using 
	\begin{equation}
		\dot{\mathbf{X}} \approx \frac{1}{T_d}(\mathbf{X}_{t+1}-\mathbf{X}_{t}),
		\label{EulerMethod}
	\end{equation}
	with $\mathbf{X}_t$ being the state of time $t$, and $T_d$ being the time interval where the device is assumed to be constant. Substituting \eqref{EulerMethod} into \eqref{basicControlModel}, the control model at time $t$ is expressed as 
	\begin{equation}  \label{disStateFuncUnorg}
		\scalebox{1}{$
			\begin{aligned}
				\frac{1}{T_d}(\mathbf{X}_{t+1}-\mathbf{X}_{t}) = \mathbf{A} \mathbf{X}_{t}+\mathbf{B} \mathbf{U}_{t}.
			\end{aligned}
			$}
	\end{equation}
	
	Suppose a linear control strategy is applied, i.e. $\mathbf{U} = \mathbf{K}\mathbf{X}$ with $\mathbf{K} = [K_1, K_2, K_3]$, which is widely applied in \cite{LinearCtrl_1,LinearCtrl_2,LinearCtrl_3}. The control model \eqref{disStateFuncUnorg} can be reorganized as
	\begin{equation}\label{reorgDisCtrlFunc}
		\begin{aligned}
			\mathbf{X}_{t+1} = \widetilde{\mathbf{A}}\mathbf{X}_{t}+\widetilde{\mathbf{B}} \mathbf{K}\mathbf{X}_t ,
		\end{aligned}
	\end{equation}
	where $\widetilde{\mathbf{A}}=T_d \mathbf{A}+\mathbf{I}$, and $\widetilde{\mathbf{B}}=T_d\mathbf{B}$.
	
	\subsubsection{Control Model Considering Imperfect Sensing And Wireless Communication}
	The imperfect sensing, followed by the subsequent estimation process, complicates the controller's task of generating control commands. 
	These commands depend on accurate state information of the actuator, and any inaccuracies in this process can lead to control deviation.
	To model the impact of the imperfect sensing and estimation on the dynamic function, let $\hat{\mathbf{X}}_t$ denote the estimate of the state $\mathbf{X}$. 
	The device then receives the estimated-state-based control command $\hat{\mathbf{U}}_t = \mathbf{K}\hat{\mathbf{X}}_t$, rather than the actual-state-based command $\mathbf{U}_t = \mathbf{K}\mathbf{X}_t$.
	Therefore, \eqref{reorgDisCtrlFunc} can be developed as
	\begin{equation}\label{sensingEffectCtrlFunc}
		\begin{aligned}
			\mathbf{X}_{t+1} = \widetilde{\mathbf{A}}\mathbf{X}_{t}+\widetilde{\mathbf{B}}\mathbf{K}\hat{\mathbf{X}}_t.
		\end{aligned}
	\end{equation}
	
\begin{figure}
		\centering
		\includegraphics[width=0.35\textheight]{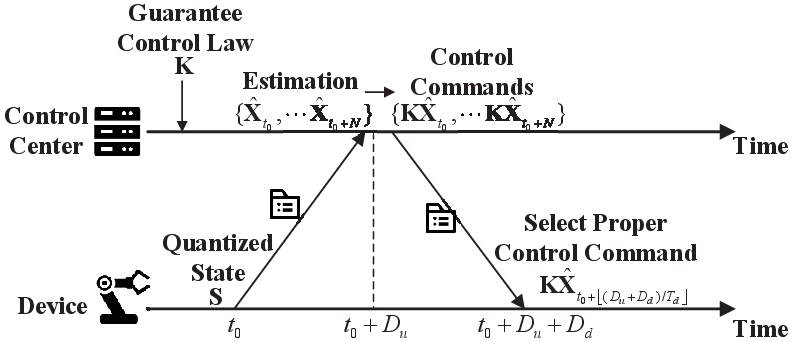}
		\DeclareGraphicsExtensions.
		\caption{The delay-compensation control method.}	
		\label{ControlScheme}
	\end{figure}

	The effect of wireless communication can be attributed to the inaccurate control caused by communication delay and the loss of control instructions due to the package loss. When the communication delay occurs, the device will receive control commands corresponding to the previous state rather than the current state, resulting in a suboptimal control strategy.
	However, the delay-compensated strategy can be applied to compensate for the impact of communication delay \cite{DelayedMPC_1, DelayedMPC_2}.
	Therefore, a delay-compensation control method is proposed, as Fig. \ref{ControlScheme}.
	Based on the control objective, the control center generates the optimal linear control law coefficient $\mathbf{K}$ in advance.
	After the current device's state is uploaded to the control center at time $t_0$, the estimated state $\hat{\mathbf{X}}_{t_0}$ is obtained through estimation, and the states for the subsequent $N$ time instants are subsequently estimated.
	Based on the estimated $N$ state values, control sequence for the corresponding $N$ time instants are generated.
	The sequence is transmitted to the device entirely, and appropriate data packets are selected based on the discretized time interval from the sending of perception data to the receiving of control packets, i.e. \textcolor{black}{$\lfloor D_c/T_d \rfloor$}, where $\lfloor\cdot \rfloor$ represents the floor function.

	Based on the above analysis, the impact of packet loss on the iteration of system states is further considered.
	When the packet loss occurs, the control command cannot be transmitted to the device, and the state equation relies on the physical properties during uncontrolled iterations, that is
	\begin{equation} \label{StateFunc}
		\scalebox{1}{$
			\begin{aligned}
				\mathbf{X}_{t+1} = \widetilde{\mathbf{A}}\mathbf{X}_{t}+ \eta\widetilde{\mathbf{B}} \mathbf{K}\hat{\mathbf{X}}_{t},
			\end{aligned}
			$}
	\end{equation}
	where $\eta$ is a random variable which equals $0$ with probability $\epsilon_c$ and equals $1$ with probability $1-\epsilon_c$.
	\subsection{Sensing and Estimation Model}
	The sensing-estimation process is located in the first half of the closed-loop process. 
	The sensor detects the current state of the actuator and quantifies it into a digital signal, which is then transmitted to the control center.  
	At the control center, the future states are estimated according to the received information to compensate for the communication delays, which is introduced as the control model and Fig. \ref{ControlScheme}.
	In the subsequent analysis, a uniform quantization method is adopted. 
	Further analysis is conducted on the errors introduced by the quantization and estimation.
	
	Suppose the sensors are capable of locally acquiring the state information of the actuators with high accuracy, i.e. the data prior to quantization is accurate.
	\textcolor{black}{Besides, the detection of the quantization is assumed to lie in the area of $[\mathbf{X}_L, \mathbf{X}_U]$.}
	Denote the quantization level is $r$, i.e. the area $[\mathbf{X}_L, \mathbf{X}_U]$ is uniformly quantized into $2^r$ intervals.
	Hence, the sensing data generated per unit time, also known as the arrival rate of the uplink queue, can be represented as 
	\begin{equation} \label{arrivUplink}
		\scalebox{1}{$
			\begin{aligned}
				\lambda_u = \frac{r}{T_d}.
			\end{aligned}
			$}
	\end{equation}
	
	Besides, the midpoint of the interval is taken as the quantization of the state $\mathbf{X}_t$, which is denoted by $\mathbf{X}_t^q$.
	\begin{equation} \label{quantize} 
		\scalebox{1}{$
			\begin{aligned}
				\mathbf{X}_t^q = \mathbf{X}_L + (j+\frac{1}{2}) \cdot \frac{1}{2^r}[\mathbf{X}_U-\mathbf{X}_L],
			\end{aligned}
			$}
	\end{equation}
	where $j \in \left\{0,1,\cdots,2^r-1\right\}$. 
	Therefore, the quantization error for $\mathbf{X}_t$ can be formulated as 
	\begin{equation} \label{estiErrorAtT}
		\scalebox{1}{$
			\begin{aligned}
				\mathbf{e}_{0} =& \mathbf{X}_t^q-\mathbf{X}_t \in \left[\frac{1}{2}\frac{1}{2^r}[\mathbf{X}_L-\mathbf{X}_U],\frac{1}{2}\frac{1}{2^r}[\mathbf{X}_U-\mathbf{X}_L]\right].
			\end{aligned}
			$}
	\end{equation}
	On the assumption that the error $\mathbf{e}_0$ follows the uniform distribution, which is a usual assumption in a generalized control system \cite{UniformEstiError}, we obtain 
	\begin{equation} 
		\scalebox{1}{$
			\begin{aligned}
				\mathbb{E}[\mathbf{e}_{0}] = \mathbb{E}[\mathbf{X}_t^q-\mathbf{X}_t] = 0,
			\end{aligned}
			$}
	\end{equation}
	\begin{equation} \label{varianceE0}
		\scalebox{1}{$
			\begin{aligned}
				\mathbb{E}[\mathbf{e}_{0}^\mathrm{T}\mathbf{e}_{0}]= \frac{1}{12} \frac{1}{4^r} [(\mathbf{X}_L-\mathbf{X}_U)^\mathrm{T}(\mathbf{X}_L-\mathbf{X}_U)].
			\end{aligned}
			$}
	\end{equation}
	
	After quantization, the estimator at the controller should estimate the future $N$ states for generating the control sequence to compensate the communication delay, as Fig. \ref{ControlScheme}.
	According to \eqref{StateFunc}, the estimation is made by
	\begin{equation} \label{estiStateFunc}
		\scalebox{1}{$
			\begin{aligned}
				\hat{\mathbf{X}}_{{t+1}} = &\widetilde{\mathbf{A}}\hat{\mathbf{X}}_{t}+\mathbb{E}(\eta)\widetilde{\mathbf{B}} \mathbf{K}\hat{\mathbf{X}}_{t}\\
				=&\widetilde{\mathbf{A}}\hat{\mathbf{X}}_{t}+(1-\epsilon_c)\widetilde{\mathbf{B}} \mathbf{K}\hat{\mathbf{X}}_{t}\\
				\dot{=} & \textbf{A}_{K} \hat{\mathbf{X}}_{t},
			\end{aligned}
			$}
	\end{equation}
	where $\textbf{A}_{K} = \widetilde{\mathbf{A}}+(1-\epsilon_c)\widetilde{\mathbf{B}}$.
	According to the iteration \eqref{estiStateFunc}, the estimation of the future state at time $t+\tau$ is
	\begin{equation} 
		\scalebox{1}{$
			\begin{aligned}
				\hat{\mathbf{X}}_{{t+\tau}} = (\textbf{A}_{K})^\tau \hat{\mathbf{X}}_{t},
			\end{aligned}
			$}
	\end{equation}
	Therefore, the mean estimation error of time $t+\tau$ is obtained as
	\begin{equation} \label{eq_mean_etau}
		\scalebox{1}{$
			\begin{aligned}
				\mathbb{E}[\mathbf{e}_\tau] = \mathbb{E}[\hat{\mathbf{X}}_{{t+\tau}} - \mathbf{X}_{t+\tau}] =(\textbf{A}_{K})^\tau (\mathbf{X}_t^q - \mathbf{X}_{t}) = 0,
			\end{aligned}
			$}
	\end{equation}
	
	Besides, when $\mathrm{Tr}(\widetilde{\mathbf{A}}^\mathrm{T}\widetilde{\mathbf{A}}) \neq 1$, the mean square estimation error satisfies
	\begin{equation} \label{ineq_variance_etau_1}
		\scalebox{1}{$
			\begin{aligned}
				&\quad\mathbb{E}[\mathbf{e}_\tau^\mathrm{T}\mathbf{e}_\tau] \\ &\le\frac{1}{12} \frac{1}{4^r}[\mathrm{Tr}(\widetilde{\mathbf{A}}^\mathrm{T}\widetilde{\mathbf{A}})]^{\tau} [(\mathbf{X}_L-\mathbf{X}_U)^\mathrm{T}(\mathbf{X}_L-\mathbf{X}_U)]\\
				&+(\epsilon_c-\epsilon_c^2)\mathrm{Tr}(\mathbf{K}^\mathrm{T}\widetilde{\mathbf{B}}^\mathrm{T}\widetilde{\mathbf{B}} \mathbf{K}) \mathbf{X}_M^\mathrm{T}\mathbf{X}_M\frac{1-[\mathrm{Tr}(\widetilde{\mathbf{A}}^\mathrm{T}\widetilde{\mathbf{A}})]^{\tau}}{1-\mathrm{Tr}(\widetilde{\mathbf{A}}^\mathrm{T}\widetilde{\mathbf{A}})}.
			\end{aligned}
			$}
	\end{equation}
	And when $\mathrm{Tr}(\widetilde{\mathbf{A}}^\mathrm{T}\widetilde{\mathbf{A}}) = 1$,
	\begin{equation} \label{ineq_variance_etau_2}
		\scalebox{1}{$
			\begin{aligned}
				&\mathbb{E}[\mathbf{e}_{\tau}^\mathrm{T}\mathbf{e}_{\tau}]\\
				\le &\frac{1}{12} \frac{1}{4^r}[(\mathbf{X}_L-\mathbf{X}_U)^\mathrm{T}(\mathbf{X}_L-\mathbf{X}_U)]
				+(\epsilon_c-\epsilon_c^2)\tau \\& \cdot\mathrm{Tr}(\mathbf{K}^\mathrm{T}\widetilde{\mathbf{B}}^\mathrm{T}\widetilde{\mathbf{B}} \mathbf{K}) \mathbf{X}_M^\mathrm{T}\mathbf{X}_M,
			\end{aligned}
			$}
	\end{equation}
	where $\mathbf{X}_M = \{x_{M,i}|x_{M,i} = \max\{|x_{U,i}|,|x_{L,i}|\}\}$\textcolor{black}{, with $x_{U,i} $ and $x_{L,i}$ being the $i$th element of matrix $\mathbf{X}_U$ and $\mathbf{X}_L$, respectively}, where $i\in \left\{1,2,\dots,n\right\}$.  The proof of \eqref{ineq_variance_etau_1} and \eqref{ineq_variance_etau_2} are provided in Appendix D.
	According to \eqref{ineq_variance_etau_1} and \eqref{ineq_variance_etau_2}, a common bound for any $\tau$ with the delay bound $D_{c,\max}$ can be obtained as \eqref{Longeq_1}.
	
	\stripsep -6pt
	{\renewcommand{\arraystretch}{1.5}
		\begin{figure*}[ht]
			\begin{equation} \label{Longeq_1}
				\scalebox{1}{$
					\begin{aligned}
						&\mathbb{E}[\mathbf{e}_{\tau}^\mathrm{T}\mathbf{e}_{\tau}] \\
						\le &\left\{\begin{array}{l}
							\frac{1}{12} \frac{1}{4^r}[\mathrm{Tr}(\widetilde{\mathbf{A}}^\mathrm{T}\widetilde{\mathbf{A}})]^{\lfloor D_{c,\max}/T_d \rfloor} [(\mathbf{X}_L-\mathbf{X}_U)^\mathrm{T}(\mathbf{X}_L-\mathbf{X}_U)]
							+(\epsilon_c-\epsilon_c^2)\mathrm{Tr}(\mathbf{K}^\mathrm{T}\widetilde{\mathbf{B}}^\mathrm{T}\widetilde{\mathbf{B}} \mathbf{K}) \mathbf{X}_M^\mathrm{T}\mathbf{X}_M\frac{1-[\mathrm{Tr}(\widetilde{\mathbf{A}}^\mathrm{T}\widetilde{\mathbf{A}})]^{\lfloor D_{c,\max}/T_d \rfloor}}{1-\mathrm{Tr}(\widetilde{\mathbf{A}}^\mathrm{T}\widetilde{\mathbf{A}})}
							\\ \hfill \text{when}\  \mathrm{Tr}(\widetilde{\mathbf{A}}^\mathrm{T}\widetilde{\mathbf{A}}) > 1\\
							
							\frac{1}{12} \frac{1}{4^r} [(\mathbf{X}_L-\mathbf{X}_U)^\mathrm{T}(\mathbf{X}_L-\mathbf{X}_U)]+(\epsilon_c-\epsilon_c^2)\frac{\mathrm{Tr}(\mathbf{K}^\mathrm{T}\widetilde{\mathbf{B}}^\mathrm{T}\widetilde{\mathbf{B}} \mathbf{K}) \mathbf{X}_M^\mathrm{T}\mathbf{X}_M }{1-\mathrm{Tr}(\widetilde{\mathbf{A}}^\mathrm{T}\widetilde{\mathbf{A}})}  \hfill \text{when}\  \mathrm{Tr}(\widetilde{\mathbf{A}}^\mathrm{T}\widetilde{\mathbf{A}}) < 1\\
							
							\frac{1}{12} \frac{1}{4^r}[(\mathbf{X}_L-\mathbf{X}_U)^\mathrm{T}(\mathbf{X}_L-\mathbf{X}_U)]
							+(\epsilon_c-\epsilon_c^2){\lfloor D_{c,\max}/T_d \rfloor} \cdot\mathrm{Tr}(\mathbf{K}^\mathrm{T}\widetilde{\mathbf{B}}^\mathrm{T}\widetilde{\mathbf{B}} \mathbf{K}) \mathbf{X}_M^\mathrm{T}\mathbf{X}_M
							\hfill \text{when}\  \mathrm{Tr}(\widetilde{\mathbf{A}}^\mathrm{T}\widetilde{\mathbf{A}}) = 1
						\end{array}\right.\\
						\dot{=} & F_e(r,\epsilon_c,D_{c,\max})
					\end{aligned}
					$}
			\end{equation} 
		\end{figure*}
	}

	\subsection{Convergence Analysis of the Closed-loop System}
	To quantitatively analyze the impact of communication, sensing, and estimation on the performance of control systems, this section introduces an analytical framework based on the Lyapunov theory. Through this framework, we netx derive inequalities that establish a relationship between the system's convergence rate and key factors such as delay, packet loss, and estimation errors. 
	
	According to the LaSalle's invariance principle \cite{Convergence, LyaTheory}, the system converges at a rate of $\rho$, if 
	\begin{equation} \label{rhoDefine}
		\scalebox{1}{$
			\begin{aligned}
				\mathbb{E}\left[V_f(\mathbf{X}_{t+1})|\mathbf{X}_{t}\right] \le \rho V_f(\mathbf{X}_{t}),
			\end{aligned}
			$}
	\end{equation}
	where $V_f(\cdot)$ is the Lyapunov function of the closed-loop system, which is defined as \cite[Ch. 4.2]{MPCBook}
	\begin{equation} \label{LyaDef}
		\scalebox{1}{$
			\begin{aligned}
				V_f(\textbf{X}_t) = \textbf{X}_t^T\textbf{P}\textbf{X}_t \dot{=}|\mathbf{X}_t|_\mathbf{P}^2,
			\end{aligned}
			$}
	\end{equation}
	with $\textbf{P}$ being a pre-defined semi-positive definite matrix.

	
	
	Based on \eqref{StateFunc}, \eqref{rhoDefine} and \eqref{LyaDef}, the impact of communication and sensing on the convergence of the closed-loop system is derived in Theorem~\ref{ConvergenceRateIneq}. 
	\begin{theorem}[Convergence Rate Inequality] \label{ConvergenceRateIneq}
		The sufficient condition of the system convergence \eqref{rhoDefine} is satisfied if
		\begin{equation} \label{rhoTheorem}
			\scalebox{1}{$
				\begin{aligned}
					\rho \ge \frac{F_\rho(W_u,W_d,\mathbf{K},\mathbf{X}_t)}{|\mathbf{X}_t|^2_\mathbf{P}},
				\end{aligned}
				$}
		\end{equation}
		where
		\begin{equation} 
			\scalebox{1}{$
				\begin{aligned}
					&F_\rho(W_u,W_d,\mathbf{K},\mathbf{X}_t) \\=&(1-\epsilon_c) \{-|\widetilde{\mathbf{A}} \mathbf{X}_t|_\mathbf{P}^2+|(\widetilde{\mathbf{A}} + \widetilde{\mathbf{B}}\mathbf{K}) \mathbf{X}_t|_P^2 \\
					&+F_e(r,\epsilon_c,D_{c,\max}) \mathrm{Tr}[|\mathbf{B}\mathbf{K}|_\mathbf{P}^2]\}
					+|\widetilde{\mathbf{A}} \mathbf{X}_t|_\mathbf{P}^2,
				\end{aligned}
				$}
		\end{equation}
		and $F_e(r,\epsilon_c,D_{c,\max})$ is given in \eqref{Longeq_1}.
	\end{theorem}
	\vspace{0.2cm}
	\begin{IEEEproof}
		See the proof in Appendix E.
	\end{IEEEproof}

	\section{The Optimization of Control Strategy and Resource Allocation}
	As illustrated in Fig. \ref{ControlScheme}, to ensure the on-demand equipment operations, it is essential to design the optimal control law, i.e. $\mathbf{K}$. 
	However, as can be inferred from the system model, the performance of communication, sensing, and control are intricately linked. 
	Therefore, to achieve the control objectives, it is necessary to simultaneously regulate the resources such as the bandwidth, quantization levels, and control strategies.
	Considering the complexity of such problem, a constrained \ac{MPC} strategy is applied to simultaneously generate the optimal control law and  the resource allocation strategy  \cite{MPCBook}.
	The control objective is adopted as the control cost for prediction over the upcoming $N$ time intervals, i.e., 
	\textcolor{black}{\begin{equation} \label{ControlCost}
			\scalebox{1}{$
				\begin{aligned}
					J(\mathbf{K},W_u,W_d) = &\sum_{t=0}^{N-1} \left[\hat{\mathbf{X}}_t^\text{T}\mathbf{Q}\hat{\mathbf{X}}_t+(\mathbf{K}\hat{\mathbf{X}}_t)^\text{T}\mathbf{R}(\mathbf{K}\hat{\mathbf{X}}_t)\right]\\&+ \frac{1}{2}\hat{\mathbf{X}}_N^\text{T}\mathbf{P}_f\hat{\mathbf{X}}_N,
				\end{aligned}
				$}
		\end{equation}
		where $N$ is the prediction horizon where the future states are considered. 
		The term $\hat{\mathbf{X}}_t^\mathrm{T} \mathbf{Q} \hat{\mathbf{X}}_t$, which can be reformed as $(\hat{\mathbf{X}}_t-\mathbf{0})^\mathrm{T} \mathbf{Q} (\hat{\mathbf{X}}_t-\mathbf{0})$, is the distances between $\hat{\mathbf{X}}_t$ and  the control objective $\hat{\mathbf{X}}=\mathbf{0}$ at time $t$.
		$(\mathbf{K}\hat{\mathbf{X}}_t)^\text{T}\mathbf{R}(\mathbf{K}\hat{\mathbf{X}}_t)$ represents the amount of energy expended during the control process.
		Both $\mathbf{Q}$ and $\mathbf{R}$ are predetermined positive semi-definite matrices that respectively reflect the weighted relationships between state changes and control energy, with their values carefully chosen to meet practical requirements.
		The summation term represents the total control cost over the next $N$ time instants.
		$\frac{1}{2}\hat{\mathbf{X}}_N^\text{T}\mathbf{P}_f\hat{\mathbf{X}}_N$ is the \textit{terminal cost} \cite{MPCBook}, the purpose of which is to compensate for the control cost, bringing it closer to the result when $N=\infty$, so as to enhance the control performance.}
	$\mathbf{P}_f$ is the solution of the discrete Racciti function \cite[Ch. 2.5]{MPCBook}, given by 
	\begin{equation} \label{Racciti}
		\scalebox{1}{$
			\begin{aligned}
				\mathbf{P}_f = &\mathbf{Q} + \widetilde{\mathbf{A}}^\text{T} \mathbf{P}_f \widetilde{\mathbf{A}} \\&- \widetilde{\mathbf{A}}^\text{T} \mathbf{P}_f \widetilde{\mathbf{B}} \left( \mathbf{R} + \widetilde{\mathbf{B}}^\text{T} \mathbf{P}_f \widetilde{\mathbf{B}} \right)^{-1} \widetilde{\mathbf{B}}^\text{T} \mathbf{P}_f \widetilde{\mathbf{A}}.
			\end{aligned}
			$}
	\end{equation}
	
	According to Theorem \ref{arrivalRateLimit} and Theorem \ref{ConvergenceRateIneq}, the state function \eqref{StateFunc}, and the bandwidth limit, the optimization problem is formulated as follows.
	\begin{subequations}\label{P1}
		\begin{alignat}{2} 
			\text{P1:} &\min\limits_{\substack{\mathbf{K},W_u, W_d}} \quad && J(\mathbf{K},W_u,W_d) \label{P1a}\\ 	
			& \quad \ \text { s.t. } \quad&&  \lambda_{u} \le \lambda_{\max} \label{P1b}\\
			&&& \rho \ge \frac{F_\rho(W_u,W_d,\mathbf{K},\hat{\mathbf{X}}_t)}{|\hat{\mathbf{X}_t}|^2_\mathbf{P}} \label{P1c}\\
			&&& 0 \le W_u + W_d \leq W_0 \label{P1d}\\
			&&& 0\le\epsilon_c\le1 \label{P1e}\\
			&&& \textcolor{black}{\hat{\mathbf{X}}_{t+1} = \textbf{A}_{K}\hat{\mathbf{X}}_t} \label{P1f}
		\end{alignat}
	\end{subequations}
	
	In the above optimization problem, constraint \eqref{P1b} and \eqref{P1c} are the solutions of Theorem \ref{arrivalRateLimit} and Theorem \ref{ConvergenceRateIneq}, respectively. Constraint \eqref{P1d} constrains the available bandwidth of the communication. Constraint \eqref{P1e} guarantees the correct range of packet loss rate. \textcolor{black}{Constraint \eqref{P1f} is the estimation rule in \eqref{estiStateFunc}}.
	
	Due to the complex structure of constraint \eqref{P1c}, this optimization problem  non-convex.
	The DE method is particularly effective for this kind of non-convex problems due to its robust global search capability, which significantly enhances the probability of finding the global optimum.
	Therefore, \ac{DE} based optimization method  is employed to seek the global optimal solution for this problem \cite{DE_1, DE_2}, as Algorithm \ref{Algorithm of DE}.
	\begin{algorithm}
		\caption{\ac{DE} based optimization method}
		\label{Algorithm of DE}
		\begin{algorithmic}[1] 
			\item[]\textbf{Input:}  System parameters such as $W_0$, $\theta_u$, $\theta_d$, $\text{SNR}_u$, $\text{SNR}_d$, etc. Algorithm parameters such as population size $N_p$, crossover probability $p_{cr}$, differential weight $F_d$, maximum iterations $N_m$, maximum counter $n$, and tolerance of convergence $tol$.
			\STATE Solve $\mathbf{P}_f$ according to \eqref{Racciti}
			\STATE Randomly initialize population with \( N_p\) individuals
			\STATE Initialize best fitness change counter: \( counter = 0 \)
			\FOR{\( i = 1 \) to \(N_m \) and \( counter < n \)}
			\FOR{each individual \( \mathbf{\Xi}_k = [K_k, W_{u,k}, W_{d,k}] \) }
			\STATE Randomly select three individuals \( \mathbf{\Xi}_{r1} \), \( \mathbf{\Xi}_{r2} \), \( \mathbf{\Xi}_{r3} \) from the population
			\STATE Compute the mutant vector: \( \mathbf{V}_k = \mathbf{\Xi}_{r1} + F_d \times (\mathbf{\Xi}_{r2} - \mathbf{\Xi}_{r3}) \)
			\STATE Let $\mathbf{H}_k = \mathbf{V}_k$ with probability $p_{cr}$ and $\mathbf{H}_k = \mathbf{\Xi}_i$ with probability $1-p_{cr}$
			\IF{ \{Constraints are not violated for $\mathbf{U}_k$ \AND $J(\mathbf{U}_k)<J(\mathbf{\Xi}_k)$ \}\OR Constraints are violated for $\mathbf{\Xi}_k$}
			\STATE Replace \( \mathbf{\Xi}_k \) with \( \mathbf{U}_k \) in the population
			\ENDIF
			\ENDFOR
			\IF{\( |\min J(\mathbf{H}_k) - \min J(\mathbf{\Xi}_k)| < tol \)}
			\STATE Increment \( counter \) by 1
			\ELSE
			\STATE Reset \( counter \) to 0
			\ENDIF
			\ENDFOR
			\STATE \textbf{Output:} The individual corresponding to the $Fitness_{current}$.
		\end{algorithmic}
	\end{algorithm}

	\section{Simulation Results}
	In this section, the motion control of an \acs{AGV} is simulated to demonstrate the effectiveness of the proposed control strategy and resource allocation method. Furthermore, we conduct an analysis to investigate the dynamic influence of parameter variations on the system. 
	For each experiment, \textcolor{black}{500} times of Monte-Carlo trials are conducted to ensure statistical reliability. 
	The setting of the parameters is shown in TABLE~\ref{simulation_parameters}.
	\begin{table}[!ht]
		\centering
		\renewcommand{\arraystretch}{1.3} 
		\caption{Simulation Parameters}
		\begin{tabularx}{\columnwidth}{|X|X|X|X|}
			\hline
			\textbf{Parameter} & \textbf{Value} & \textbf{Parameter} & \textbf{Value} \\
			\hline
			\multicolumn{4}{|c|}{System Parameters}\\
			\hline
			$T_d$ & \SI{0.1}{s} & $\varsigma$ & 0.125 \\
			\hline
			$\mathbf{P}$ & $\text{diag}(10,10,1)$ & $\mathbf{Q}$ & $\text{diag}(10,10,1)$ \\
			\hline
			$\mathbf{R}$ & 1 &$\mathbf{X}_0$ & $[-100,1,1]^\text{T}$\\
			\hline
			$N$ & 10 & $c_d$ & 0.1 \\
			\hline
			$\text{SNR}_u$ & \SI{30}{dB} & $\text{SNR}_d$ & \SI{33}{dB} \\
			\hline
			$\beta_u$ & 1 & $\beta_d$ & 1 \\
			\hline
			$\theta_u$ & $0.02$ & $\theta_d$ & $0.04$ \\
			\hline
			$W_0$ & \SI{1.5}{MHz} &&\\
			\hline
			\multicolumn{4}{|c|}{Algorithm Parameters}\\
			\hline
			$N_p$ & 15 & $p_{cr}$ & 0.7 \\
			\hline
			$n$ & 5 & $N_m$ & 1000 \\
			\hline
			$tol$ & 0.01 & $F_d$ & 0.5 \\
			\hline
		\end{tabularx}
		\label{simulation_parameters}
	\end{table}
	
	The iteration of the system is carried out based on the delay-compensation control method proposed in Section~\ref{ControlModel}, as shown in Fig. \ref{ControlScheme}.
	The performance of the system is evaluated by three time-varying indicators, i.e. \textit{state distance}, \textit{accumulated control energy}, and \textit{accumulated cost}. The definitions of the three indicators are provided in the sequel: 
	\begin{itemize}
		\item 
		The \textit{state distance} is the quadratic form of the state $\mathbf{X}_t$, which reveals the distance from state $\mathbf{X}_t$ to state $\mathbf{0}$, i.e.,
		\begin{equation} 
			\scalebox{1}{$
				\begin{aligned}
					state\ distance = \mathbf{X}_t^\text{T}\mathbf{Q}\mathbf{X}_t.
				\end{aligned}
				$}
		\end{equation}
		\item The \textit{accumulated control energy} is the total of control energy from the initial time to the current time, i.e., 
		\begin{equation} 
			\scalebox{1}{$
				\begin{aligned}
					acc.\ control\ energy = \sum_{t=0}^{N-1} \left[(\mathbf{K}\mathbf{X}_t)^\text{T}\mathbf{R}(\mathbf{K}\mathbf{X}_t)\right].
				\end{aligned}
				$}
		\end{equation}
		\item The \textit{accumulated cost} is the total control cost from the initial time to the current, which is the sum part of \eqref{ControlCost}, i.e., 
		\begin{equation} 
			\scalebox{1}{$
				\begin{aligned}
					acc.\ cost = \sum_{t=0}^{N-1} \left[\mathbf{X}_t^\text{T}\mathbf{Q}\mathbf{X}_t+(\mathbf{K}\mathbf{X}_t)^\text{T}\mathbf{R}(\mathbf{K}\mathbf{X}_t)\right].
				\end{aligned}
				$}
		\end{equation}
	\end{itemize}
	
	\begin{figure}[!t]
		\centering
		\vspace{-1mm}
		\subfloat[State distance v.s. closed-loop delay constraints]{\includegraphics[width=0.9\linewidth]{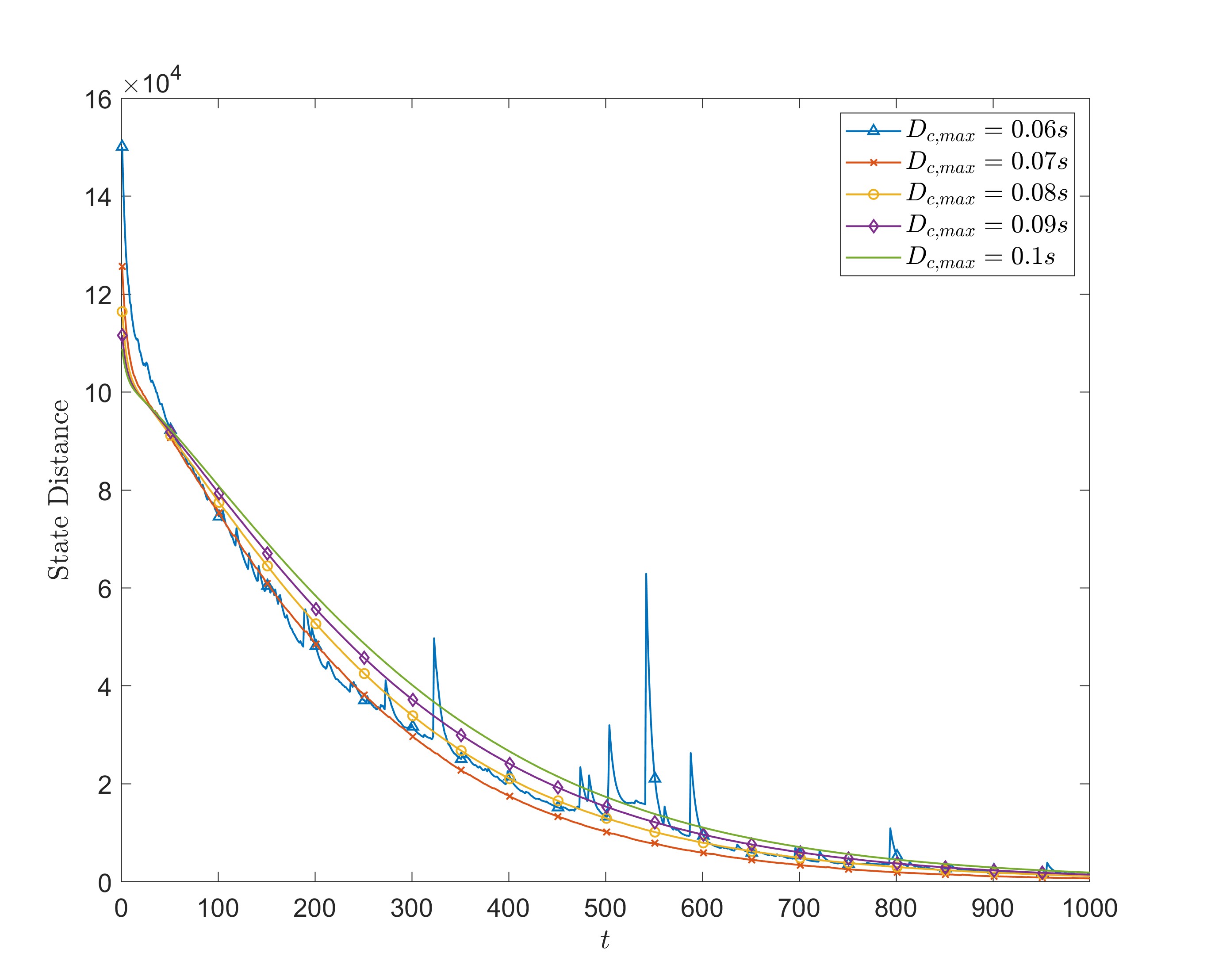}%
			\label{StateDistance_dc}}
		\vspace{-1mm}
		\subfloat[Accumulated control energy v.s. closed-loop delay constraints]{\includegraphics[width=0.9\linewidth]{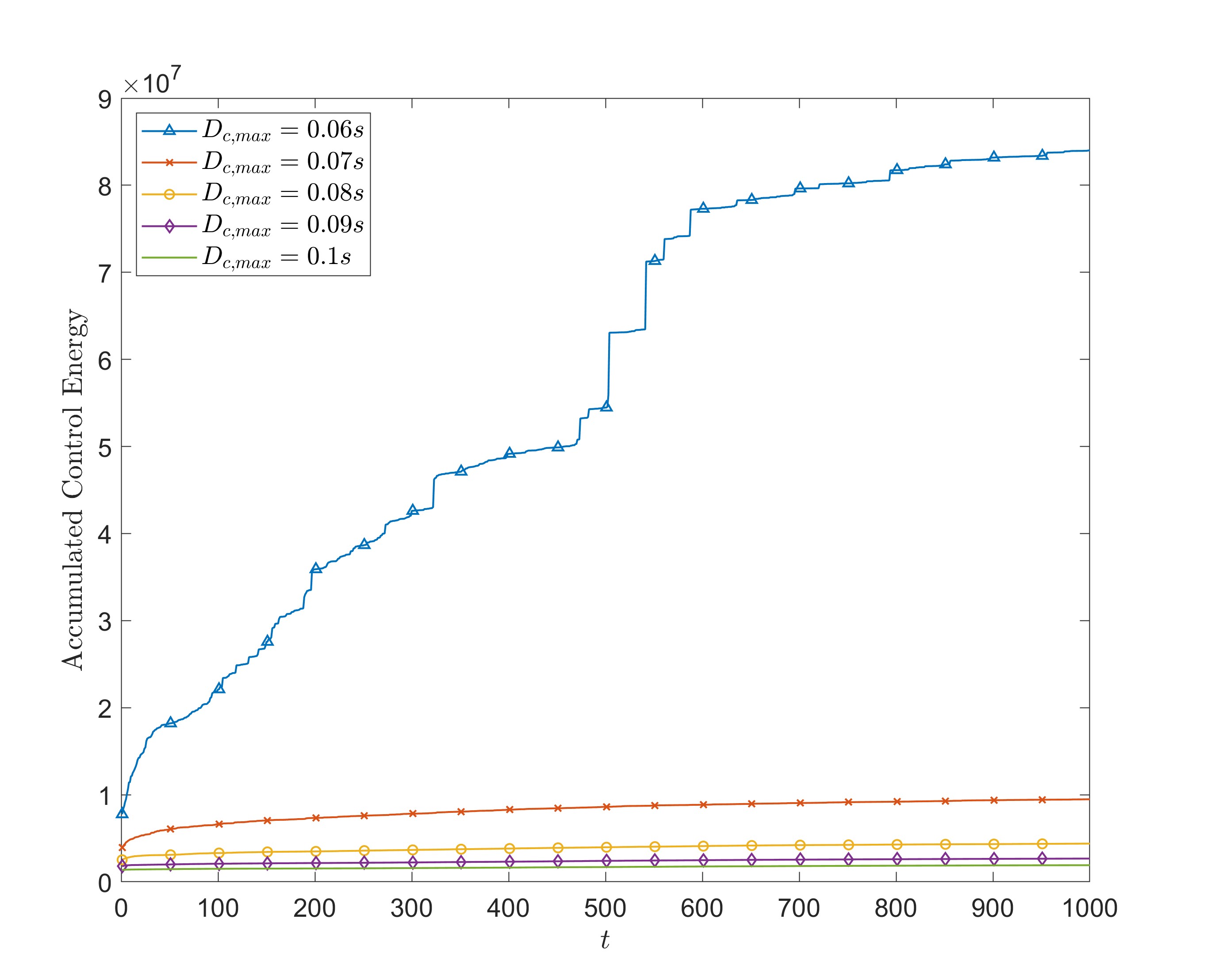}%
			\label{AccumulatedCtrlEnergy_dc}}
		\vspace{-1mm}
		\subfloat[Accumulated cost v.s. closed-loop delay constraints]{\includegraphics[width=0.9\linewidth]{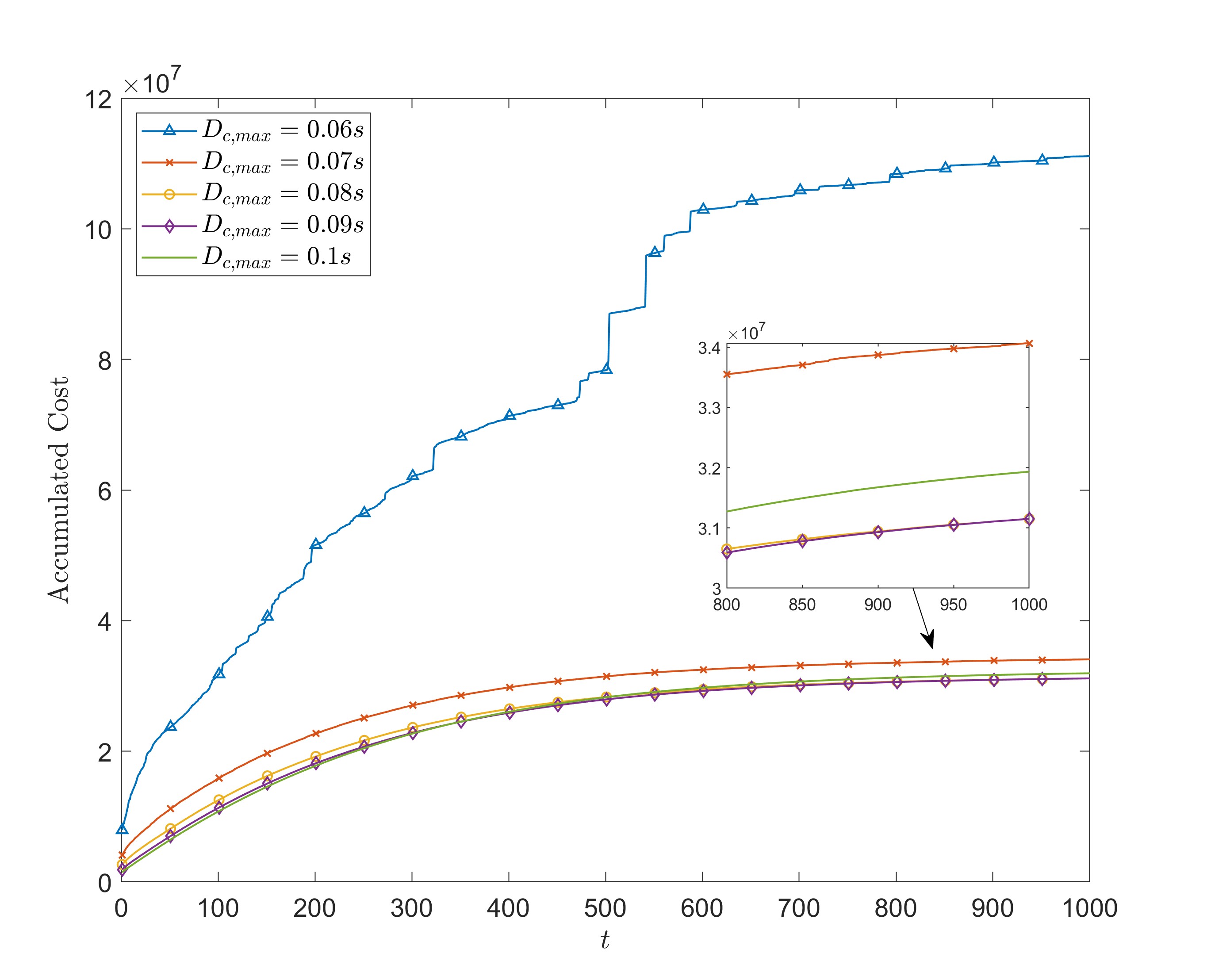}%
			\label{AccumulatedCost_dc}}
		\caption{Variations of the control cost, state distance, accumulated cost and accumulated control energy with closed-loop delay constraints.}
		\label{Dc_sim}
		\vspace{-4mm}
	\end{figure}

	\subsection{The Impact of Closed-Loop Delay}
	The impact of closed-loop delay $D_{c,\max}$ on the system is shown in Fig.~\ref{Dc_sim}\textcolor{black}{, where $D_{c,\max}$ varies from \SI{0.06}{s} to \SI{0.1}{s}, $\rho = 0.999$, and $r = 6$.}
	As illustrated in Fig.~\ref{Dc_sim}(a), it can be observed that the convergence speed of states significantly increases with the reduction of the maximum allowable closed-loop delay, which is due to the fact that strict delay constraints enhance the timeliness of control commands.
	However, when $D_{c,\max}$ is too small, such as $D_{c,\max} = \SI{0.06}{s}$ in Fig.~\ref{Dc_sim}(a), the system experiences a high packet loss rate due to \eqref{epsilon_cResult}. 
	This leads to the failure of most control commands in reaching the actuators, consequently resulting in significant fluctuations in the device's state.
	For the same reason, Fig.~\ref{Dc_sim}(b) demonstrates that excessively strict constraints on delay, \textcolor{black}{i.e. $D_{c,\max} = \SI{0.06}{s}$}, significantly augment the consumption of control energy.
	Furthermore, increasing $D_{c,\max}$ reduces the packet loss rate, diminishing the need for large control actions against packet loss impacts. Consequently, the accumulated control energy gradually decreases.
	
	From Fig.~\ref{Dc_sim}(a) and Fig.~\ref{Dc_sim}(b), it can be observed that although an increase in $D_{c,\max}$ leads to a reduction in control energy consumption, it also results in a slower convergence. 
	\textcolor{black}{
		This results in a reduction of the accumulated cost with the increase of $D_{c,\max}$ during the early stages of the control process, i.e. when $t\le500$. 
		Conversely, when $t\ge500$, within the interval where $D_{c,\max}\ge\SI{0.08}{s}$, the accumulated cost increases with the rise of $D_{c,\max}$.
		This is because that although a higher $D_{c,\max}$ entails lower accumulated control energy consumption, its slower convergence leads to a longer accumulation period for control costs. 
		Over time, this results in a higher accumulated cost.}
	
	\begin{figure}[!t]
		\centering
		\vspace{-1mm}
		\subfloat[State distance v.s. convergence rate]{\includegraphics[width=0.9\linewidth]{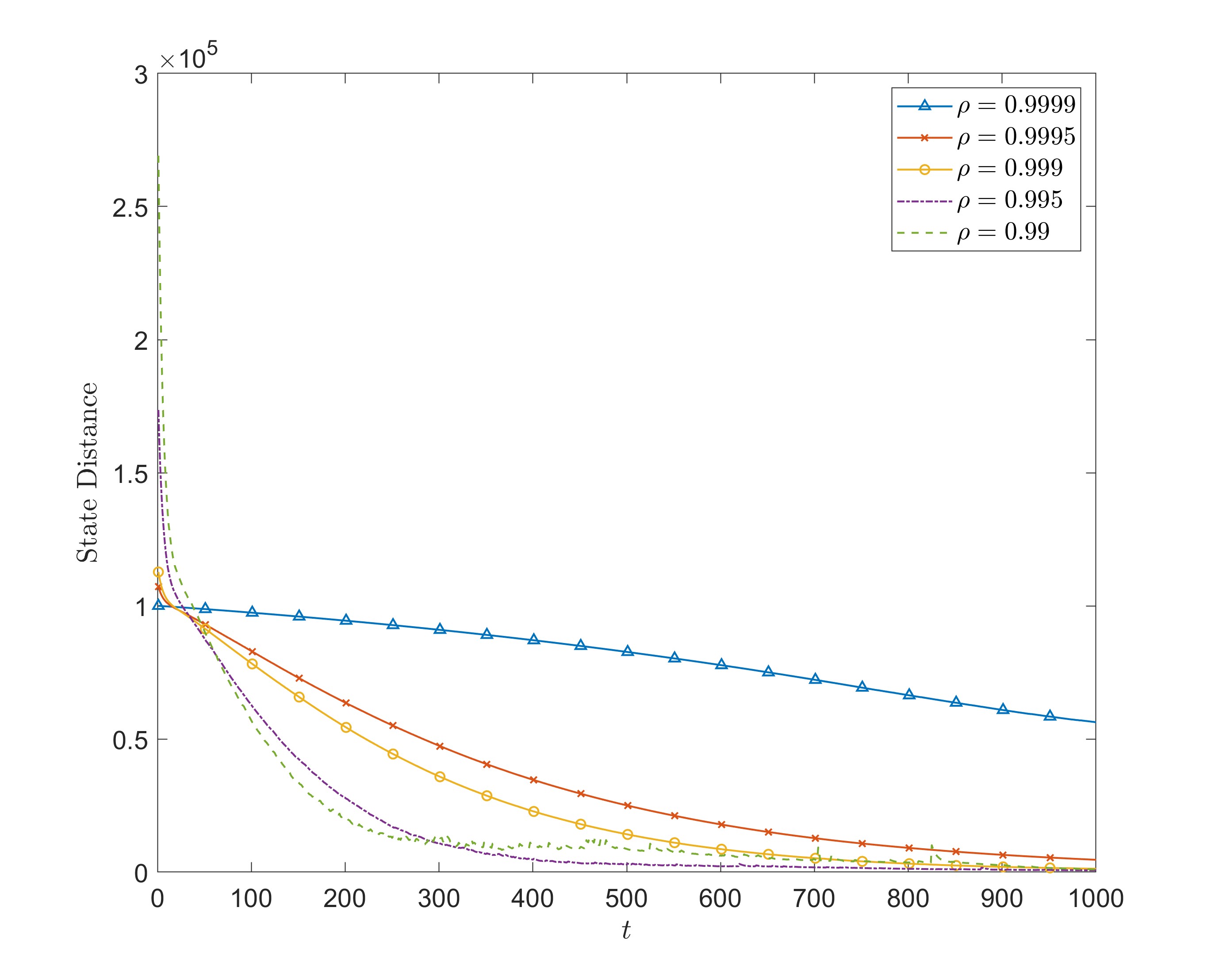}%
			\label{StateDistance_rho}}
		\vspace{-1mm}
		\subfloat[Accumulated control energy v.s. convergence rate]{\includegraphics[width=0.9\linewidth]{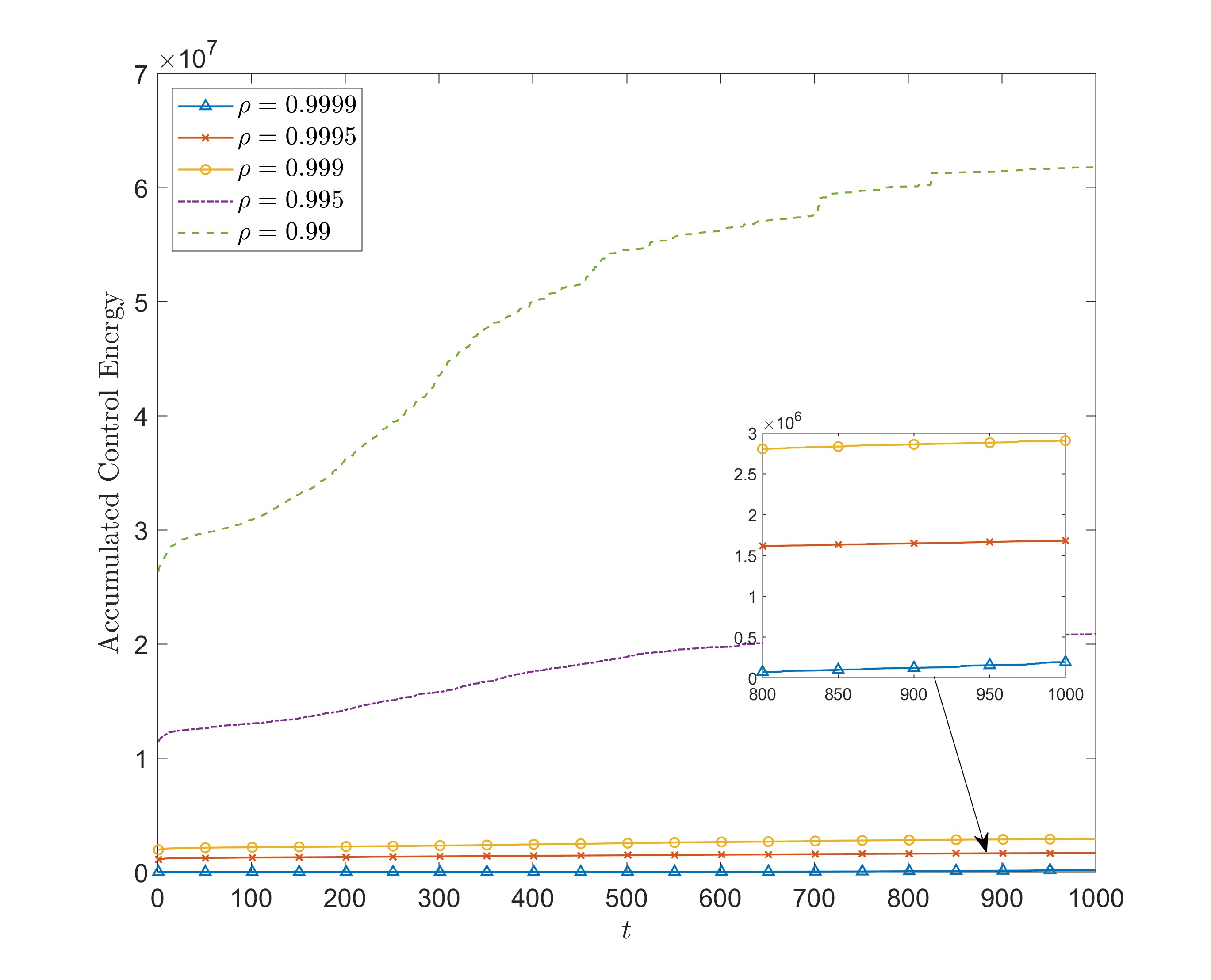}%
			\label{AccumulatedCtrlEnergy_rho}}
		\vspace{-1mm}
		\subfloat[Accumulated cost v.s. convergence rate]{\includegraphics[width=0.9\linewidth]{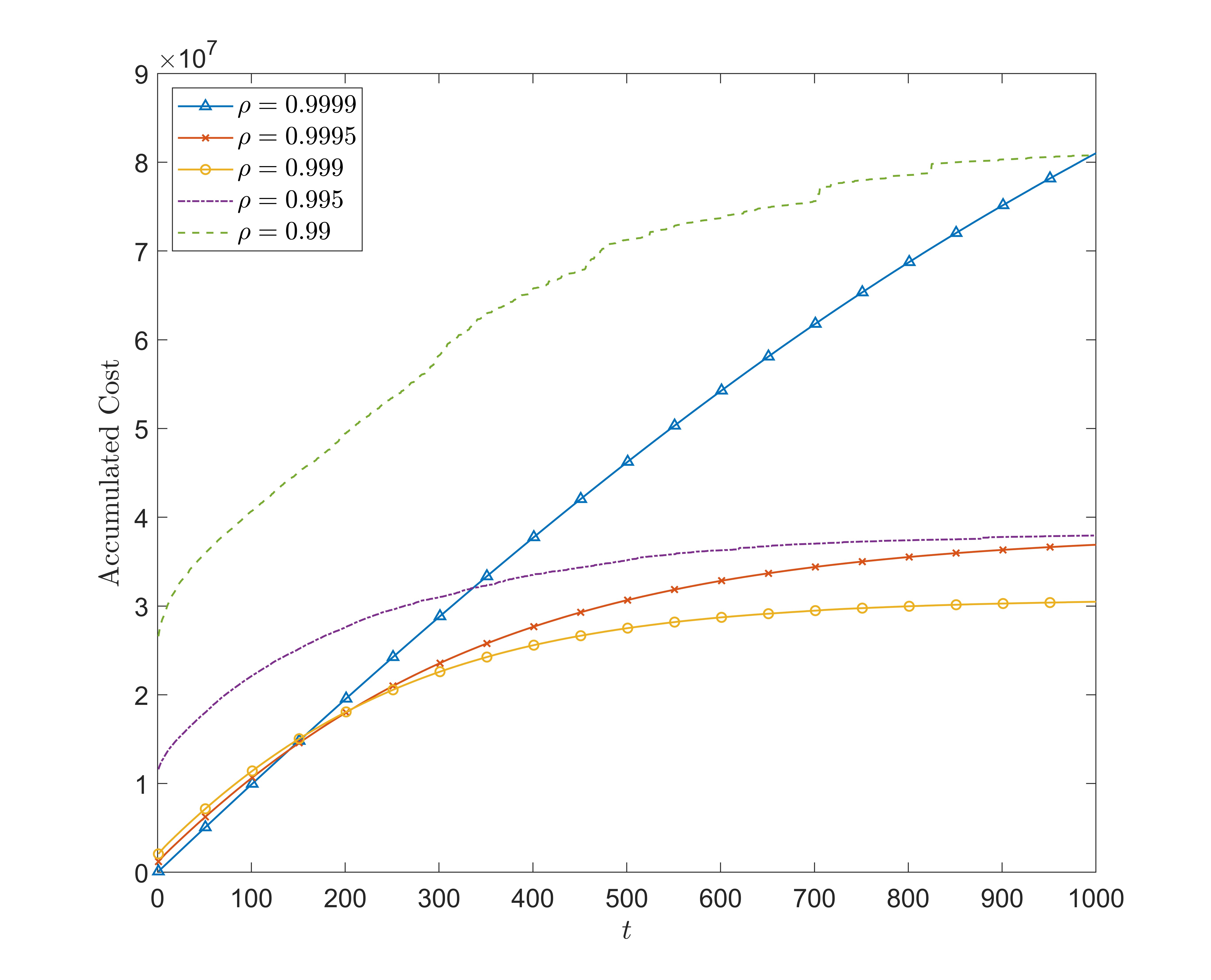}%
			\label{AccumulatedCost_rho}}
		\caption{Variations of the control cost, state distance, accumulated cost and accumulated control energy with different convergence rate.}
		\label{Rho_sim}
		\vspace{-4mm}
	\end{figure}
	\begin{figure}[!t]
		\centering
		\vspace{-1mm}
		\subfloat[State distance v.s.quantization level]{\includegraphics[width=0.9\linewidth]{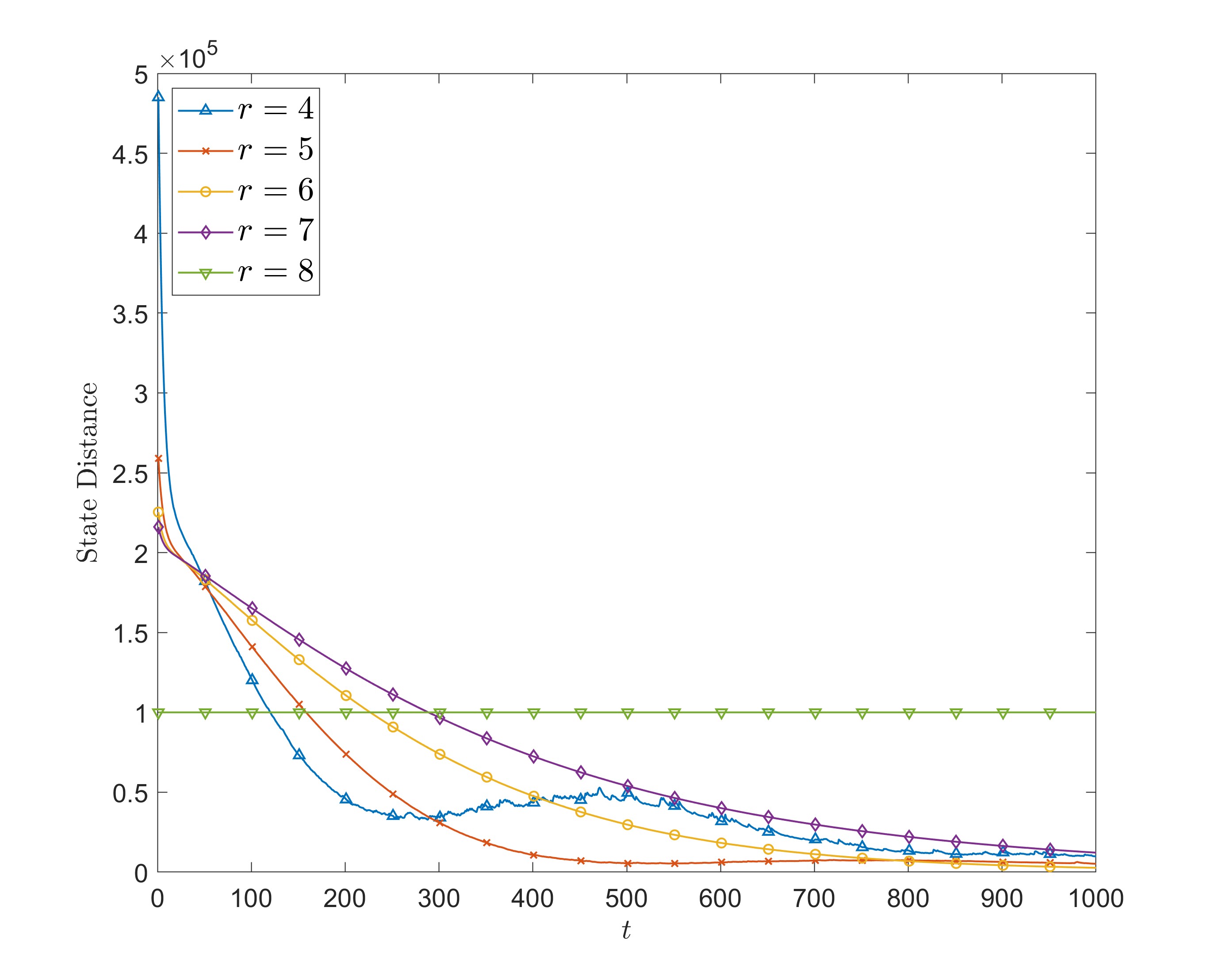}%
			\label{StateDistance_r}}
		\vspace{-1mm}
		\subfloat[Accumulated control energy v.s. quantization level]{\includegraphics[width=0.9\linewidth]{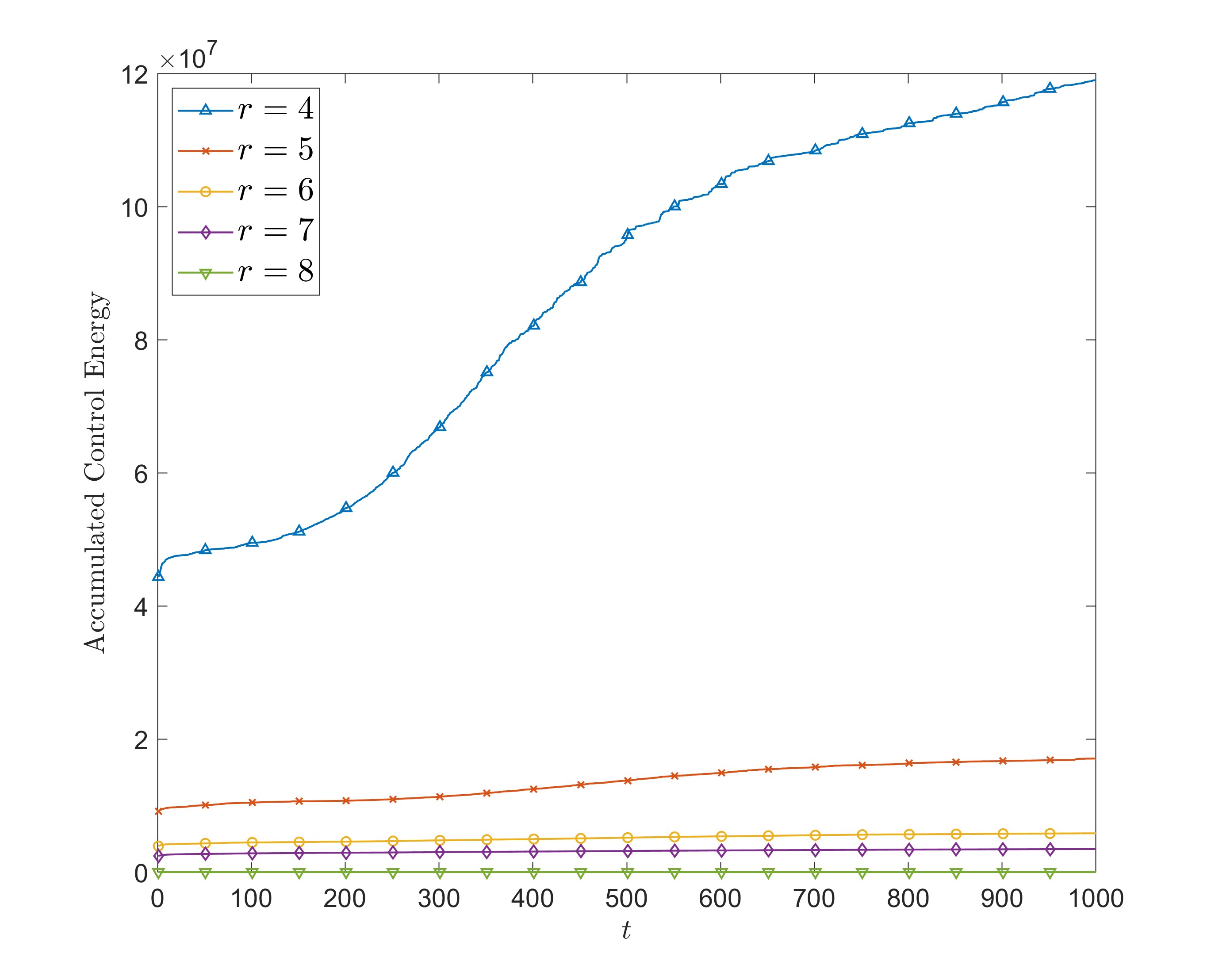}%
			\label{AccumulatedCtrlEnergy_r}}
		\vspace{-1mm}
		\subfloat[Accumulated cost v.s. quantization level]{\includegraphics[width=0.9\linewidth]{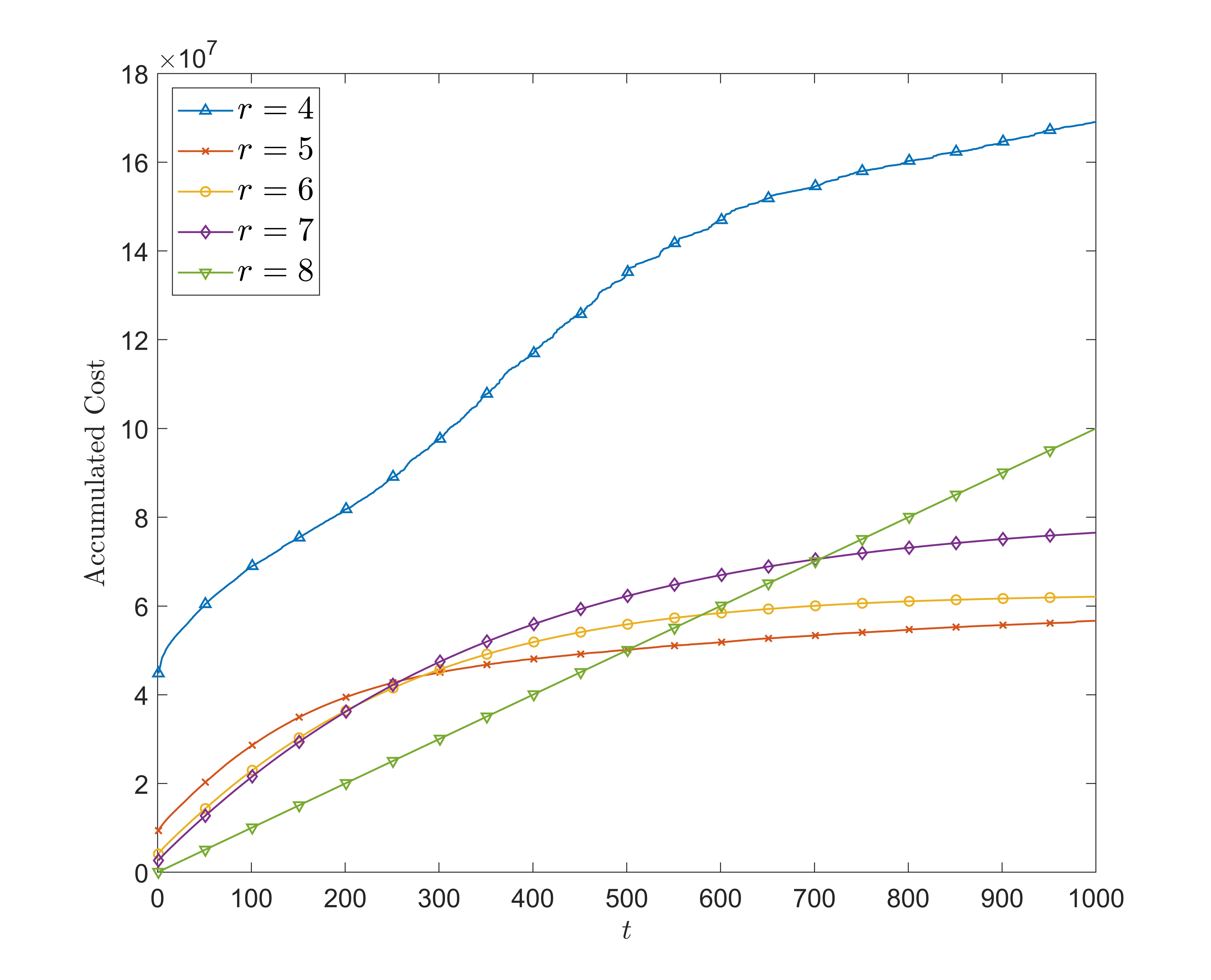}%
			\label{AccumulatedCost_r}}
		\caption{Variations of the control cost, state distance, accumulated cost and accumulated control energy with different quantization level.}
		\label{R_sim}
		\vspace{-4mm}
	\end{figure}

	\subsection{The Impact of Convergence Rate}
	The impact of the convergence rate $\rho$ on the system is shown in Fig.~\ref{Rho_sim}\textcolor{black}{, where $\rho$ varies from $0.9999$ to $0.99$, $D_{c,\max} = \SI{0.1}{s}$, and $r = 6$.}
	As shown in Fig.~\ref{Rho_sim}(a) and Fig.~\ref{Rho_sim}(b), the system cost and the state distance indeed converge faster with the decrease of $\rho$.
	However, according to \eqref{rhoTheorem}, an excessively small $\rho$, such as $\rho=0.99$ in Fig.~\ref{Rho_sim}, leads to  an increase in the packet loss rate, thereby causing greater fluctuations in the state.
	\textcolor{black}{As Fig.~\ref{Rho_sim}(c), the steady-state accumulated cost initially exhibits a downward trend with the decrease from $\rho = 0.9999$ to $\rho = 0.999$, subsequently followed by an increase with the decrease from $\rho = 0.999$ to $\rho = 0.99$.}
	This is because, on the one hand, a reduction in $\rho$ leads to an increase in the control energy consumption. On the other hand, the acceleration of convergence results in a more rapid decrease in  control energy.
	In the interval of $\rho=0.9999$ to $\rho=0.999$, the impact of convergence acceleration predominates, leading to a reduction to the  steady-state accumulated control energy with the decrease of $\rho$.
	In the interval of $\rho=0.999$ to $\rho=0.99$, the excessive increase in control energy dominates, which results in a tiny effect of the convergence acceleration.

	\subsection{The Impact of Quantization Level}
	The impact of quantization level $r$ on the system is shown in Fig.~\ref{R_sim}\textcolor{black}{, where $r$ varies from $4$ to $8$, $D_{c,\max} = \SI{0.1}{s}$, and $\rho = 0.999$.}.
	It is evident that when the value of $r$ is sufficiently small, particularly when $r=4$, the convergence of the system is hindered, leading to a substantial increase in both accumulated cost and control energy.
	This is because the estimation error is much too large, making it difficult for the control instructions to accurately identify the current state.
	Furthermore, an excessively large $r$, such as $r=8$ as depicted in the figure, leads to an unsolvable optimization problem. 
	This situation  obstructs the generation of control commands by the control center, thus limiting the actuator's capacity to achieve convergence.

	Except for the above situations, a larger $r$ results in a faster convergence and smaller control energy, as $r$ respectively equals $5$, $6$ and $7$ in Fig.~\ref{R_sim}(a) and Fig.~\ref{R_sim}(b).
	In such cases, a decrease in quantization level leads to an increase in sensing errors, subsequently requiring higher control energy from the device to sustain control operations. Ultimately, this contributes to increased control power consumption. 
	However, increasing the quantization order places a heavier burden on the communication link, resulting in an increased average time delay and slower convergence.
	Furthermore, the accumulated cost is the combination of the state distance and the accumulated control energy. 
	\textcolor{black}{Despite the higher initial control energy consumption associated with a smaller quantization level before $t=250$, a faster convergence rate is achieved for smaller $r$ when $r\ge 5$ and $t\ge 250$, which results in a lower accumulated cost overall, as demonstrated in Fig.~\ref{R_sim}(c).}

	\section{Conclusion}
	
	\textcolor{black}{The existing theoretical models for closed-loop control systems with wireless networks fail to achieve the expected control effects in practical applications, thus resulting in challenges in the design of communication, sensing and control systems.}
	To address the issue of the separation between uplink and downlink models, an uplink-downlink coupled closed-loop communication model was first proposed based on the transmission characteristics of the industrial wireless networks. 
	The closed-form expressions of closed-loop delay and packet loss rate were next derived to provide more practical metrics for the industrial system.
	Subsequently, a control model based on the delay compensation algorithm and packet loss was presented, as well as a sensing-estimation model based on quantization and estimation error.
	In order to model the relationship between communication, sensing and control, the inequality related to communication, sensing, and control parameters was derived based on the proposed models. The result provided a lower bound on the convergence rate that ensures system stability.
	Finally, to provide guidance for the design of the closed-loop system, a joint optimization problem for control and resource allocation was introduced. An algorithm based on \ac{DE} was presented to obtain the global optimal solution of the non-convex problem.
	Numerical simulations indicated that due to the interdependencies among communication, sensing, and control systems, excessively high or low parameter designs can markedly degrade the system's control effectiveness.

	\section{Appendix A \\ Proof of Theorem 1}
	Substituting \eqref{Ri} into \eqref{EC}, there is 
	\begin{equation}\label{ECwithException}
		\scalebox{1}{$
			\begin{aligned}
				&C_i(\theta_i, W_i, \text{SNR}_i,\beta_i)\\
				=&-\frac{1}{\theta_i}\ln \left(\mathbb{E}\left\{\exp 
				\left[{-\theta_i W_i \log _2\left(1+\text{SNR}_i \gamma^2\right)}\right]\right\}\right).
			\end{aligned}
			$}
	\end{equation}
	
	Considering that $\gamma^2$ follows an exponential distribution with parameter $\beta_i$, \eqref{ECwithException} can be further derived as
	\begin{equation}\label{expectationEC}
		\scalebox{1}{$
			\begin{aligned}
				&C_i(\theta_i, W_i, \text{SNR}_i,\beta_i)\\=&W_i \log _2(\text{SNR}_i \beta_i)-\frac{1}{\text{SNR}_i \beta_i}\\
				&-\frac{1}{\theta_i} \ln \left(\Gamma\left(1-\frac{W_i \theta_i}{\ln 2}, \frac{1}{\text{SNR}_i \beta_i}\right)\right) ,
			\end{aligned}
			$}
	\end{equation}
	where $\Gamma(s,x)$ is the upper incomplete gamma function which is defined as
	\begin{equation}
		\Gamma(s, x)=\int_x^{\infty} t^{s-1} e^{-t} d t.
	\end{equation}
	
	Besides, $1-\frac{W_i \theta_i}{\ln 2} <0$ and $\frac{1}{\text{SNR}_i \beta_i} \rightarrow 0$ since the order of magnitude for bandwidth $W_i$ is typically above $10^6$, the order of $\text{SNR}_i$ is usually above $10^3$, $\theta$ is around $10^{-3}$ for factory scenario, and the typical value for $\beta_i$ is approximately $1$.
	Subsequently, 
	\begin{equation}\label{gammaFunc}
		\scalebox{1}{$
			\begin{aligned}
				&\Gamma\left(1-\frac{W_i \theta_i}{\ln 2}, \frac{1}{\text{SNR}_i \beta_i}\right)\\ =&
				\left(\frac{1}{\text{SNR}_i \beta_i}\right)^{\left(1-\frac{W_i \theta_i}{\ln 2}\right)}E_{\frac{W_i \theta_i}{\ln 2}}\left(\frac{1}{\text{SNR}_i \beta_i}\right)\\
				&\approx \left(\frac{1}{\text{SNR}_i \beta_i}\right)^{\left(1-\frac{W_i \theta_i}{\ln 2}\right)} \cdot \frac{-1}{\left(1-\frac{W_i \theta_i}{\ln 2}\right)},
			\end{aligned}
			$}
	\end{equation}
	where $E_p(x)$ is the generalized exponential integral function, which is
	\begin{equation} 
		\scalebox{1}{$
			\begin{aligned}
				E_p(x)=\int_1^{\infty} \frac{\mathrm{e}^{-x t}}{t^p} \mathrm{~d} t,
			\end{aligned}
			$}
	\end{equation} 
	and the approximation holds due to 
	\begin{equation}
		E_p(0) = \frac{1}{1-p}.
	\end{equation}
	
	Substituting \eqref{gammaFunc} into \eqref{expectationEC}, 
	\begin{equation}
		\scalebox{1}{$
			\begin{aligned}
				&C_i(\theta_i, W_i, \text{SNR}_i,\beta_i)\\
				=&W_i \log _2(\text{SNR}_i \beta_i)-\frac{1}{\text{SNR}_i \beta_i}\\
				&+ {1 \over \theta_i }\left(1 - {{W_i\theta_i } \over {\ln 2}}\right)\ln \left(\text{SNR}_i\beta_i\right) \\
				&+ {1 \over \theta_i }\ln \left({{W_i\theta_i } \over {\ln 2}} - 1\right)\\
				=&\frac{1}{\theta_i} \ln (\text{SNR}_i \beta_i)(\frac{W_i \theta_i}{\ln 2}-1)-\frac{1}{\text{SNR}_i \beta_i}.
			\end{aligned}
			$}
	\end{equation}
	
	\section{Appendix B\\ Proof of Theorem 2}
	According to \eqref{singleQueueDelay}, the delay of a packet in a single queue follows a exponential distribution with parameter $\theta_i C_i(\theta_i, W_i, \text{SNR}_i,\beta_i)$.
	Therefore, the probability density function of $D_i(i=u,d)$ is
	\begin{equation} 
		\scalebox{1}{$
			\begin{aligned}
				f_{D_i}(x) = \mu_i \exp(-\mu_i x)\ \ (i=u,d),
			\end{aligned}
			$}
	\end{equation}
	where according to Theorem \ref{ECSingleQueue},
	\begin{equation} 
		\scalebox{1}{$
			\begin{aligned}
				\mu_i =& \theta_i C_i(\theta_i, W_i, \text{SNR}_i,\beta_i)\\
				=& \ln (\text{SNR}_i \beta_i)(\frac{W_i \theta_i}{\ln 2}-1)-\frac{\theta_i}{\text{SNR}_i \beta_i}\ \ (i=u,d).
			\end{aligned}
			$}
	\end{equation}
	
	Thus, the probability density function the closed-loop delay $D_c = D_u+D_d$ is the convolution of the probability density functions of $D_u$ and $D_d$, that is 
	\begin{equation} 
		\scalebox{1}{$
			\begin{aligned}
				f_{D_c}(x) =& f_{D_u}(x) \ast f_{D_d}(x)\\
				=&-\frac{\left(\mathrm{e}^{-x \mu_u}-\mathrm{e}^{-x \mu_d}\right) \mu_u \mu_d}{\mu_u-\mu_d}.
			\end{aligned}
			$}
	\end{equation}
	
	Subsequently, the package drop probability $\epsilon_c$ satisfies
	\begin{equation} 
		\scalebox{1}{$
			\begin{aligned}
				\begin{aligned}
					\epsilon_c=& \mathrm{P}\left\{D_c>D_{c, \max }\right\}  =1-\int_0^{D_{c, \max }} f_{D_c}(x) \mathrm{d} x \\
					=&\frac{\mathrm{e}^{-D_{c, \operatorname{mxx}} \mu_d} \mu_u-\mathrm{e}^{-D_{c, \max } \mu_u} \mu_d}{\mu_u-\mu_d}.
				\end{aligned}
			\end{aligned}
			$}
	\end{equation}
	
	Besides, the exception of the closed-loop delay when package is not dropped is
	\begin{equation} 
		\scalebox{1}{$
			\begin{aligned}
				\begin{aligned}
					& E\left[D_c|D_c<D_{c,\max}\right] \\ =&\int_0^{D_{c, \text { max }}}  \frac{xf_{D_c}(x)}{P(D_c<D_{c,\max})} \mathrm{d} x \\
					=&\frac{1}{\mu_u \mu_d\left(\mu_u-\mu_d+\mathrm{e}^{-D_{c, \max x} \mu_u} \mu_d-\mathrm{e}^{-D_{c, \max } \mu_d} \mu_u\right)}\\
					&\cdot [\mu_u^2-\mu_d^2+\mathrm{e}^{-D_{c, \max } \mu_u} \mu_d^2\left(1+D_{c, \max } \mu_u\right)\\
					&\ \ -\mathrm{e}^{-D_{c, \max x} \mu_d} \mu_u^2\left(1+D_{c, \max } \mu_d\right)].
				\end{aligned}
			\end{aligned}
			$}
	\end{equation}

	\section{Appendix C\\ Proof of Theorem 3}
	According to \cite[Eqn. 10]{TandemQ}, the departure process of the first queue $L_u$ is
	\begin{equation}  \label{DepRateUp}
		\scalebox{1}{$
			\begin{aligned}
				L_u=\left\{\begin{array}{l}
					\lambda_{u}, \quad 0 \leq \theta_d \leq \theta_u \\
					\frac{1}{\theta_d}\{\left(\theta_d-\theta_u\right) C_{u}(\theta_u, W_u, \text{SNR}_u,\beta_u)\\
					+\lambda_{u} \theta_u\}, \quad \theta_d>\theta_u
				\end{array}\right. .
			\end{aligned}
			$}
	\end{equation}
	
	Therefore, substituting \eqref{DepRateUp} into \eqref{lambdaD},
	\begin{equation}  \label{lambdaDLong}
		\scalebox{1}{$
			\begin{aligned}
				\lambda_d = \left\{\begin{array}{l}
					c_d\lambda_{u}, \quad 0 \leq \theta_d \leq \theta_u \\
					c_d\frac{1}{\theta_d}\{\left(\theta_d-\theta_u\right) C_{u}(\theta_u, W_u, \text{SNR}_u,\beta_u)\\
					+\lambda_{u} \theta_u\}, \quad \theta_d>\theta_u
				\end{array}\right. .
			\end{aligned}
			$}
	\end{equation}
	
	Then substituting \eqref{lambdaDLong} into \eqref{EAleEC}, when $\theta_u > \theta_d$,
	\begin{equation} \label{thetauLarger}
		\scalebox{1}{$
			\begin{aligned}
				C_d(\theta_d, W_d, \text{SNR}_d,\beta_d) \ge c_d \lambda_u,
			\end{aligned}
			$}
	\end{equation}
	and when $\theta_u < \theta_d$,
	\begin{equation} \label{thetadLarger}
		\scalebox{1}{$
			\begin{aligned}
				&C_d(\theta_d, W_d, \text{SNR}_d,\beta_d) \\ \ge &c_d\frac{1}{\theta_d}\{\left(\theta_d-\theta_u\right) 
				C_{u}(\theta_u, W_u, \text{SNR}_u,\beta_u)
				+\lambda_{u} \theta_u\}.
			\end{aligned}
			$}
	\end{equation}
	After rearranging formulas \eqref{thetauLarger} and \eqref{thetadLarger}, Theorem \ref{arrivalRateLimit} can be proved.
	
	\section{Appendix D \\ Proof of equation \eqref{ineq_variance_etau_1} and \eqref{ineq_variance_etau_2}}
	Substitute \eqref{StateFunc} and \eqref{estiStateFunc} into the left side of \eqref{ineq_variance_etau_1}, we can get
	\begin{equation} \label{Ee1Appendix1}
		\scalebox{1}{$
			\begin{aligned}
				&\mathbb{E}[\mathbf{e}_{\tau}^\mathrm{T}\mathbf{e}_{\tau}] \\
				=&\mathbb{E}[(\hat{\mathbf{X}}_{t+\tau} - \mathbf{X}_{t+\tau})^\mathrm{T}(\hat{\mathbf{X}}_{t+\tau} - \mathbf{X}_{t+\tau})]\\
				=&\mathbb{E}\{[\widetilde{\mathbf{A}}(\hat{\mathbf{X}}_{t+\tau}-\mathbf{X}_{t+\tau})+(1-\epsilon_c-\eta)\widetilde{\mathbf{B}} \mathbf{K}\hat{\mathbf{X}}_{t+\tau}]^\mathrm{T}\\
				&\quad [\widetilde{\mathbf{A}}(\hat{\mathbf{X}}_{t+\tau}-\mathbf{X}_{t+\tau})+(1-\epsilon_c-\eta)\widetilde{\mathbf{B}} \mathbf{K}\hat{\mathbf{X}}_{t+\tau}] \}\\
				=&  \mathbb{E}\{\mathbf{e}_{\tau-1}^\mathrm{T}(\widetilde{\mathbf{A}}^\mathrm{T}\widetilde{\mathbf{A}})\mathbf{e}_{\tau-1}\}+\mathbb{E}[(1-\epsilon_c-\eta)^2\hat{\mathbf{X}}_{t+\tau-1}^\mathrm{T}\mathbf{K}^\mathrm{T}\\
				&\widetilde{\mathbf{B}}^\mathrm{T}\widetilde{\mathbf{B}} \mathbf{K}\hat{\mathbf{X}}_{t+\tau-1} ]\\
				= & \mathbb{E}\{\mathrm{Tr}[(\widetilde{\mathbf{A}}^\mathrm{T}\widetilde{\mathbf{A}})\mathbf{e}_{\tau-1}\mathbf{e}_{\tau-1}^\mathrm{T}]\}+\mathbb{E}[(1-\epsilon_c-\eta)^2\\
				&\cdot\mathrm{Tr}(\mathbf{K}^\mathrm{T}\widetilde{\mathbf{B}}^\mathrm{T}\widetilde{\mathbf{B}} \mathbf{K}\hat{\mathbf{X}}_{t+\tau-1}\hat{\mathbf{X}}_{t+\tau-1}^\mathrm{T}) ].
			\end{aligned}
			$}
	\end{equation}
	
	Since $\mathrm{Tr}(\mathbf{X}\mathbf{Y})\le \mathrm{Tr}(\mathbf{X})\mathrm{Tr}(\mathbf{Y})$
	if $\mathbf{X}$ and $\mathbf{Y}$ are semidefinite matrices, \eqref{Ee1Appendix1} can be further scaled as
	\begin{equation} 
		\scalebox{1}{$
			\begin{aligned}
				&\mathbb{E}[\mathbf{e}_{\tau}^\mathrm{T}\mathbf{e}_{\tau}]\\
				\le&\mathrm{Tr}(\widetilde{\mathbf{A}}^\mathrm{T}\widetilde{\mathbf{A}})\mathbb{E}[\mathbf{e}_{\tau-1}^\mathrm{T}\mathbf{e}_{\tau-1}]+\mathbb{E}[(1-\epsilon_c-\eta)^2]\\
				&\cdot\mathrm{Tr}(\mathbf{K}^\mathrm{T}\widetilde{\mathbf{B}}^\mathrm{T}\widetilde{\mathbf{B}} \mathbf{K})\mathbb{E}[\hat{\mathbf{X}}_{t+\tau-1}^\mathrm{T}\hat{\mathbf{X}}_{t+\tau-1}]\\
				\le& \mathrm{Tr}(\widetilde{\mathbf{A}}^\mathrm{T}\widetilde{\mathbf{A}})\mathbb{E}[\mathbf{e}_{\tau-1}^\mathrm{T}\mathbf{e}_{\tau-1}]+(\epsilon_c-\epsilon_c^2)\mathrm{Tr}(\mathbf{K}^\mathrm{T}\widetilde{\mathbf{B}}^\mathrm{T}\widetilde{\mathbf{B}} \mathbf{K})\\
				&\cdot\mathbf{X}_M^\mathrm{T}\mathbf{X}_M,
			\end{aligned}
			$}
	\end{equation}
	where the first inequality is due to the Cauchy-Schwarz inequality.
	
	According to such recurrence relationship, the upper bound of $\mathbb{E}[\mathbf{e}_{\tau}^\mathrm{T}\mathbf{e}_{\tau}]$ can be obtained with $\mathbb{E}[\mathbf{e}_{0}^\mathrm{T}\mathbf{e}_{0}]$ given by \eqref{varianceE0}, i.e. when $\mathrm{Tr}(\widetilde{\mathbf{A}}^\mathrm{T}\widetilde{\mathbf{A}}) \neq 1$,
	\begin{equation} 
		\scalebox{0.97}{$
			\begin{aligned}
				&\mathbb{E}[\mathbf{e}_{\tau}^\mathrm{T}\mathbf{e}_{\tau}] \\
				\le & [\mathrm{Tr}(\widetilde{\mathbf{A}}^\mathrm{T}\widetilde{\mathbf{A}})]^{\tau}\mathbb{E}[\mathbf{e}_{0}^\mathrm{T}\mathbf{e}_{0}]+(\epsilon_c-\epsilon_c^2)\mathrm{Tr}(\mathbf{K}^\mathrm{T}\widetilde{\mathbf{B}}^\mathrm{T}\widetilde{\mathbf{B}} \mathbf{K})\\
				&\mathbf{X}_M^\mathrm{T}\mathbf{X}_M\frac{1-[\mathrm{Tr}(\widetilde{\mathbf{A}}^\mathrm{T}\widetilde{\mathbf{A}})]^{\tau}}{1-\mathrm{Tr}(\widetilde{\mathbf{A}}^\mathrm{T}\widetilde{\mathbf{A}})}\\
				=&\frac{1}{12} \frac{1}{4^r}[\mathrm{Tr}(\widetilde{\mathbf{A}}^\mathrm{T}\widetilde{\mathbf{A}})]^{\tau} [(\mathbf{X}_L-\mathbf{X}_U)^\mathrm{T}(\mathbf{X}_L-\mathbf{X}_U)]\\
				&+(\epsilon_c-\epsilon_c^2)\mathrm{Tr}(\mathbf{K}^\mathrm{T}\widetilde{\mathbf{B}}^\mathrm{T}\widetilde{\mathbf{B}} \mathbf{K}) \mathbf{X}_M^\mathrm{T}\mathbf{X}_M\frac{1-[\mathrm{Tr}(\widetilde{\mathbf{A}}^\mathrm{T}\widetilde{\mathbf{A}})]^{\tau}}{1-\mathrm{Tr}(\widetilde{\mathbf{A}}^\mathrm{T}\widetilde{\mathbf{A}})},
			\end{aligned}
			$}
	\end{equation}
	and when $\mathrm{Tr}(\widetilde{\mathbf{A}}^\mathrm{T}\widetilde{\mathbf{A}}) = 1$
	\begin{equation} 
		\scalebox{1}{$
			\begin{aligned}
				&\mathbb{E}[\mathbf{e}_{\tau}^\mathrm{T}\mathbf{e}_{\tau}]\\
				\le &\frac{1}{12} \frac{1}{4^r}[(\mathbf{X}_L-\mathbf{X}_U)^\mathrm{T}(\mathbf{X}_L-\mathbf{X}_U)]
				+(\epsilon_c-\epsilon_c^2)\tau \\& \cdot\mathrm{Tr}(\mathbf{K}^\mathrm{T}\widetilde{\mathbf{B}}^\mathrm{T}\widetilde{\mathbf{B}} \mathbf{K}) \mathbf{X}_M^\mathrm{T}\mathbf{X}_M.
			\end{aligned}
			$}
	\end{equation}

	\section{Appendix E\\ Proof of Theorem 4}
	According to \eqref{StateFunc} derived in the control model, the Lyapunov function of the closed-loop system considering the effect of sensing and communication is derived as follows.

	\begin{equation} 
		\scalebox{1}{$
			\begin{aligned}
				&\mathbb{E}[V_f(\mathbf{X}_{t+1})|\mathbf{X}_{t}] \\
				= & \mathbb{E}\left[\textbf{X}_{t+1}^T\textbf{P}\textbf{X}_{t+1}|\mathbf{X}_{t}\right]\\
				=& \mathbb{E}\left[(\widetilde{\mathbf{A}}\mathbf{X}_{t}+ \eta\widetilde{\mathbf{B}} \mathbf{K}\hat{\mathbf{X}}_{t})^\text{T}\mathbf{P}(\widetilde{\mathbf{A}}\mathbf{X}_{t}+ \eta\widetilde{\mathbf{B}} \mathbf{K}\hat{\mathbf{X}}_{t})\right]\\
				=& \epsilon_c |\widetilde{\mathbf{A}} \mathbf{X}_t|_\mathbf{P}^2  +(1-\epsilon_c)|(\widetilde{\mathbf{A}} + \widetilde{\mathbf{B}}\mathbf{K}) X_t|_P^2\\
				&+ (1-\epsilon_c) \mathbb{E}_\tau[\mathbf{e}_\tau]^\text{T}\mathbf{K}^\text{T}\mathbf{B}^\text{T}\mathbf{P}\mathbf{X}_t\\
				&+(1-\epsilon_c)\mathbf{X}^\text{T}_t\mathbf{P}\mathbf{B}\mathbf{K}\mathbb{E}_\tau[\mathbf{e}_\tau]\\
				&+(1-\epsilon_c) \mathbb{E}_\tau[\mathbf{e}_\tau^\text{T}\mathbf{K}^\text{T}\mathbf{B}^\text{T}\mathbf{P}\mathbf{B}\mathbf{K}\mathbf{e}_\tau],
			\end{aligned}
			$}
	\end{equation}
	where $\mathbb{E}_\tau[\cdot]$ represents the exception of the random variable $\tau$.
	
	According to \eqref{eq_mean_etau},
	\begin{equation} 
		\scalebox{0.95}{$
			\begin{aligned}
				\mathbb{E}_\tau[\mathbf{e}_\tau] = 0.
			\end{aligned}
			$}
	\end{equation}
	
	Therefore, considering that with the Cauchy-Schwarz inequality,
	\begin{equation} 
		\scalebox{1}{$
			\begin{aligned}
				&\mathbb{E}_\tau[\mathbf{e}_\tau^\text{T}\mathbf{K}^\text{T}\mathbf{B}^\text{T}\mathbf{P}\mathbf{B}\mathbf{K}\mathbf{e}_\tau] \\= &\mathbb{E}_\tau[\mathrm{Tr}\{\mathbf{e}_\tau^\text{T}\mathbf{K}^\text{T}\mathbf{B}^\text{T}\mathbf{P}\mathbf{B}\mathbf{K}\mathbf{e}_\tau\}]\\
				=& \mathbb{E}_\tau[\mathrm{Tr}\{\mathbf{e}_\tau\mathbf{e}_\tau^\text{T}\mathbf{K}^\text{T}\mathbf{B}^\text{T}\mathbf{P}\mathbf{B}\mathbf{K}]\}\\
				\le & \mathbb{E}_\tau[\mathrm{Tr}\{\mathbf{e}_\tau\mathbf{e}_\tau^\text{T}\}]\mathrm{Tr}\{\mathbf{K}^\text{T}\mathbf{B}^\text{T}\mathbf{P}\mathbf{B}\mathbf{K}\}\\
				=& \mathrm{Tr}\{\mathbb{E}[
				\mathbf{e}_\tau\mathbf{e}_\tau^\text{T}]\}\mathrm{Tr}\{\mathbf{K}^\text{T}\mathbf{B}^\text{T}\mathbf{P}\mathbf{B}\mathbf{K}\}\\
				=& \mathbb{E}_\tau[
				\mathbf{e}_\tau^\text{T}\mathbf{e}_\tau]\mathrm{Tr}\{\mathbf{K}^\text{T}\mathbf{B}^\text{T}\mathbf{P}\mathbf{B}\mathbf{K}\}.
			\end{aligned}
			$}
	\end{equation}
	
	Besides, according to \eqref{LongFormula},
	\begin{equation} 
		\scalebox{1}{$
			\begin{aligned}
				\mathbb{E}_\tau[
				\mathbf{e}_\tau^\text{T}\mathbf{e}_\tau]
				\le F_e(r,\epsilon_c,D_{c,\max}).
			\end{aligned}
			$}
	\end{equation}
	
	Therefore, 
	\begin{equation} \label{EVf}
		\scalebox{1}{$
			\begin{aligned}
				&\mathbb{E}\left[V_f(\mathbf{X}_{t+1})|\mathbf{X}_{t}\right] \\
				\le &\epsilon_c |\widetilde{\mathbf{A}} \mathbf{X}_t|_\mathbf{P}^2  +(1-\epsilon_c)|(\widetilde{\mathbf{A}} + \widetilde{\mathbf{B}}\mathbf{K}) X_t|_P^2\\
				&+(1-\epsilon_c) F_e(r,\epsilon_c,D_{c,\max})\operatorname{Tr}\{\mathbf{K}^\text{T}\mathbf{B}^\text{T}\mathbf{P}\mathbf{B}\mathbf{K}\}\\
				=&\epsilon_c |\widetilde{\mathbf{A}} \mathbf{X}_t|_\mathbf{P}^2  +(1-\epsilon_c)|(\widetilde{\mathbf{A}} + \widetilde{\mathbf{B}}\mathbf{K}) X_t|_P^2\\
				&+(1-\epsilon_c)F_e(r,\epsilon_c,D_{c,\max})\mathrm{Tr}\{\mathbf{K}^\text{T}\mathbf{B}^\text{T}\mathbf{P}\mathbf{B}\mathbf{K}\}.
			\end{aligned}
			$}
	\end{equation}
	
	Substituting \eqref{epsilon_cResult} into \eqref{EVf}, and 
	let \eqref{EVf} less than $\rho V_f(\mathbf{X}_{t})$, we can get the sufficient condition for \eqref{rhoDefine}, that is
	\begin{equation} \label{LongFormula}
		\scalebox{1}{$
			\begin{aligned}
				&\epsilon_c |\widetilde{\mathbf{A}} \mathbf{X}_t|_\mathbf{P}^2+(1-\epsilon_c)|(\widetilde{\mathbf{A}} + \widetilde{\mathbf{B}}\mathbf{K}) X_t|_P^2\\
				&+(1-\epsilon_c)F_e(r,\epsilon_c,D_{c,\max})\mathrm{Tr}\{|\mathbf{B}\mathbf{K}|_\mathbf{P}^2\}\\
				&\le \rho |\mathbf{X}_t|^2_\mathbf{P},
			\end{aligned}
			$}
	\end{equation}
	where $\mathbb{E}[
	\mathbf{e}_\tau^\text{T}\mathbf{e}_\tau]$ is given by \eqref{ineq_variance_etau_1} and \eqref{ineq_variance_etau_2}.
	
	After rearranging \eqref{LongFormula}, \eqref{rhoDefine} can be obtained.
	
	\begin{acronym}
		\acro{NCS}[NCS]{networked control system}
		\acro{SNR}[SNR]{singal-to-noise ratio}
		\acro{AGV}[AGV]{automated guided vehicle}
		\acro{MPC}[MPC]{model predictive control}
		\acro{DE}[DE]{differential evolution}
		\acro{LQR}[LQR]{linear quadratic regulator}
	\end{acronym}

\end{document}